\documentclass[reprint,aps,superscriptaddress,amsmath,amssymb,longbibliography]{revtex4-1}
\pdfoutput=1  
\usepackage{graphicx}
\usepackage{dcolumn}
\usepackage{bm}
          
\usepackage{color}
\usepackage{xspace}
\usepackage{ulem}
\usepackage[colorlinks=true,urlcolor=blue,citecolor=blue,linkcolor=blue,bookmarks=false, pdfstartview={FitH}]{hyperref}
\begin{document}
    
\title{Superconductivity and electronic fluctuations in Ba$_{1-x}$K$_{x}$Fe$_2$As$_2$ studied by Raman scattering} 
   
\author{S.-F. Wu}
\affiliation{Department of Physics and Astronomy, Rutgers University, Piscataway, NJ 08854, USA}
\affiliation{Beijing National Laboratory for Condensed Matter Physics, and Institute of Physics, Chinese Academy of Sciences, Beijing 100190, China}
\author{P. Richard}\email{p.richard@iphy.ac.cn}
\affiliation{Beijing National Laboratory for Condensed Matter Physics, and Institute of Physics, Chinese Academy of Sciences, Beijing 100190, China}
\affiliation{School of Physical Sciences, University of Chinese Academy of Sciences, Beijing 100190, China}
\affiliation{Collaborative Innovation Center of Quantum Matter, Beijing, China}
\author{H. Ding}
\affiliation{Beijing National Laboratory for Condensed Matter Physics, and Institute of Physics, Chinese Academy of Sciences, Beijing 100190, China}
\affiliation{School of Physical Sciences, University of Chinese Academy of Sciences, Beijing 100190, China}
\affiliation{Collaborative Innovation Center of Quantum Matter, Beijing, China}
\author{H.-H. Wen}
\affiliation{National Laboratory of Solid State Microstructures and Department of Physics, Nanjing University, Nanjing 210093, China}
\affiliation{Collaborative Innovation Center of Advanced Microstructures, Nanjing University, China}
\author{Guotai Tan}
\affiliation{Department of Physics and Astronomy, Rice University, Houston, TX 77005, USA}
\author{Meng Wang}
\affiliation{Department of Physics, University of California, Berkeley, California 94720, USA }
\author{Chenglin Zhang}
\affiliation{Department of Physics and Astronomy, Rice University, Houston, TX 77005, USA}
\author{Pengcheng Dai}
\affiliation{Department of Physics and Astronomy, Rice University, Houston, TX 77005, USA}
\author{G. Blumberg}\email{girsh@physics.rutgers.edu}
\affiliation{Department of Physics and Astronomy, Rutgers University, Piscataway, NJ 08854, USA}
\affiliation{National Institute of Chemical Physics and Biophysics, 12618 Tallinn, Estonia}
\date{\today}

\begin{abstract}

Using polarization-resolved electronic Raman  scattering we study under-doped, optimally-doped and over-doped  Ba$_{1-x}$K$_{x}$Fe$_2$As$_2$ samples  in the  normal  and  superconducting states. We show that low-energy nematic fluctuations are universal for all studied doping range. In  the  superconducting state,  we observe two distinct superconducting pair breaking  peaks  corresponding   to  one  large  and  one small superconducting gaps. In addition, we detect  a collective mode below the superconducting transition in the B$_{2g}$ channel and determine the evolution of its binding energy with doping.  Possible scenarios are proposed to explain  the  origin of the in-gap collective mode. In the superconducting state  of the  under-doped regime, we detect a re-entrance transition below which the spectral  background  changes and the collective mode vanishes.
\end{abstract} 



%
\date{\today}
\maketitle
\section{Introduction}
Multi-band systems often exhibit complex phase diagrams. Host to spin-density-wave and nematic order in the underdoped regime and critical behavior for dopings near the maximum superconducting (SC) transition temperature $T_c$, the Fe-based superconductors provide a play-ground for studying many-body electronic interactions and emerging collective modes. 
Although still debated, many theories claim that the unconventional superconductivity of the Fe-based superconductors itself derives from effective low-energy electronic interactions \cite{MazinPhysicaC2009, Graser_NJP2009,Fernandes_Chubukov_review}, thus justifying the quest for a thorough understanding of their nature.   
  
For the high-$T_c$ cuprate superconductors, one of the hallmarks of unconventional superconductivity was the observation of a neutron spin resonance mode appearing in the SC state at the antiferromagnetic wave vector $\mathbf{Q}$ \cite{Rossat-Mignod_PhysicaC185,Mook_PRL70,Fong_Nature398,Dai_Nature406,Dai_Science295,Eschrig_AdvPhys55,Zhao_PRL99,Fong1995PRL,Blumberg1995PRB}. 
Interestingly, a similar magnetic resonance mode has also been detected at 14 meV in the archetype Fe-based superconductor Ba$_{0.6}$K$_{0.4}$Fe$_2$As$_2$ \cite{christianson2008NatureINS,ZhangCL_SciRep1}. Corresponding signatures of bosonic modes were also detected by single electron spectroscopies such as angle-resolved photoemission spectroscopy (ARPES) \cite{Richard2009PRLkink} and scanning tunneling spectroscopy (STS) \cite{shan2012PRLSTM}. A sharp mode at 10~meV has also been reported in the parent compound \cite{Wu_PRB82}. These observations confirm the existence of collective excitations in the Fe-based superconductors. However, due to the complex coupling between the spin, charge, lattice and orbital degrees of freedom \cite{Fernandes2014NatPhy}, their origin is more difficult to interpret than for the simpler single band cuprates.

For the Fe-based superconductors, electronic Raman spectroscopy, which directly couples to spin singlet charge excitation at zero momentum, has recently revealed in-gap collective modes which have never been reported for the cuprates or conventional superconductors. For example, strong and sharp in-gap modes were observed for the NaFe$_{1-x}$Co$_x$As (the Na-111 electron-doped family) superconductors in both the fully-symmetric and the quadrupolar channels\,\cite{Thorsmolle2016PRB}. In-gap Raman active modes were also reported for the electron-doped Ba(Fe$_{1-x}$Co$_x$)$_2$As$_2$ family\,\cite{Gallais2016PRL} and for hole-doped Ba$_{0.6}$K$_{0.4}$Fe$_2$As$_2$\,\cite{Rudi_2013PRL,Rudi_2014PRX,bohm2016PSSB}. While several interpretations for these remarkable resonances were proposed \cite{BS1961PR,Tsuneto1960PR,Klein1984,Klein_PRB82,LeeWC2009PRL,Maiti2015PRB,scalapino2009PRB,Maiti2016arxiv,Gallais2016PRL,Thorsmolle2016PRB,Chubukov2009PRB,Khodas2014PRB,Chubukov2016PRB}, the origin of the electronic interactions leading to these in-gap resonances for multi-band Fe-based superconductors remains unresolved and calls for more extensive studies.

In this work we use polarization-resolved Raman spectroscopy to study the Ba$_{1-x}$K$_{x}$Fe$_2$As$_2$ family of superconductors as function of the hole-doping, in both the normal and SC states. We demonstrate that the critical quadrupolar nematic charge fluctuations of XY-symmetry persist across the entire phase diagram, similar to the family of electron-doped materials\,\cite{Thorsmolle2016PRB}. In addition, nematic fluctuations of (X$^2$-Y$^2$)-symmetry have also been detected. In the SC state, we observe pair-breaking coherence peaks at energies consistent with the values reported by single-particle spectroscopies. In addition, we study the evolution of the binding energy of the XY-symmetry in-gap collective mode with doping. We report a re-entrance behavior from the four-fold symmetry broken to the four-fold symmetry preserved phase in the SC state of the underdoped Ba$_{0.75}$K$_{0.25}$Fe$_2$As$_2$.

In Sec.~\ref{Experiment}, we introduce the sample preparation and the Raman experiments. We present our Raman results for three dopings in the A$_{1g}$, B${_{1g}}$ and B$_{2g}$ symmetry channels in Sec.~\ref{Normal state} and Sec.~\ref{Superconducting state} for the normal and SC states, respectively. In Sec.~\ref{Discussion}, we discuss possible scenarios for the origin of the in-gap mode. The results are summarized in Sec.~\ref{Conclusions}.

\section{Experiment}\label{Experiment}
Single crystals of Ba$_{1-x}$K$_{x}$Fe$_2$As$_2$ (x = 0.25, 0.4 and 0.6, with $T_c$ values of 31\,K, 38\,K and 25\,K, respectively) were grown by the self-flux method as described in Ref.\,\cite{2011HHWen_PRB84}. These samples are labeled UD (under-doped), OPD (optimally-doped) and OD (over-doped), respectively. The crystals used for Raman scattering were cleaved in nitrogen gas atmosphere and positioned in a continuous flow liquid helium optical cryostat. Since the optimally-doped sample was cleaved twice, the corresponding sets of data are labeled ``OPD\#1" and ``OPD\#2". 

The measurements presented here were performed in a quasi-back scattering geometry along the $c$-axis using a Kr$^+$ ion laser. Except for inset of of Fig.~\ref{Fig5_UD}(c), for which the 752\,nm (1.65 eV) laser line was used, all data were recorded with 647.1 nm (1.92 eV) excitation. The incident laser beam was focused onto a $50\times100$ $\mu$m$^2$ spot on the $ab$-surface, with an incident power smaller than 10 and 3 mW for measurements in the normal and SC states, respectively. The scattered light was collected and analyzed by a triple-stage Raman spectrometer designed for high-stray light rejection and throughput, and recorded using a liquid nitrogen-cooled charge-coupled detector. The Raman spectra were corrected for the spectral responses of the spectrometer and detector. The temperature has been corrected for the laser heating. 

\begin{figure}[!t]
\begin{center}
\includegraphics[width=3.4in]{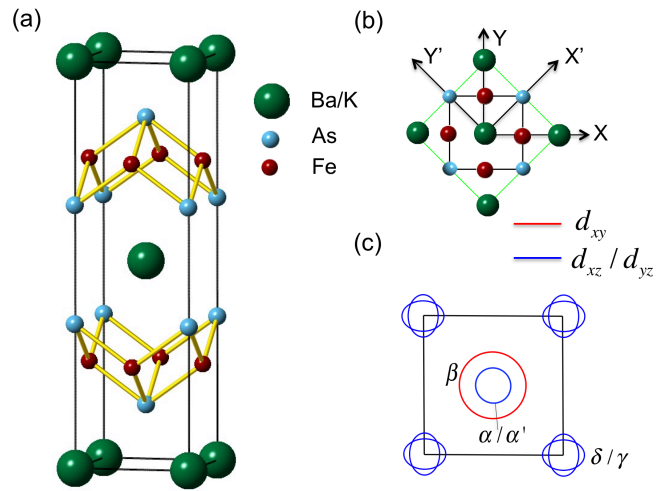}
\end{center}
\caption{\label{Fig1_FS}(Color online) (a) Crystal structure of Ba$_{1-x}$K$_{x}$Fe$_2$As$_2$. (b) Definition of the X, Y, X$^{\prime}$ and Y$^{\prime}$ directions. The green and black lines represent the 4-Fe and 2-Fe unit cells, respectively. (c) Schematic representation of the Fermi surface of Ba$_{1-x}$K$_{x}$Fe$_2$As$_2$ in the 2-Fe Brillouin zone.}
\end{figure}

In this manuscript, we define X and Y along the 2~Fe unit cell crystallographic axes $a$ and $b$ (at 45$^{\circ}$ degrees from the Fe-Fe direction) in the tetragonal phase, whereas X$'$ and Y$'$ are along the Fe-Fe directions, as shown is Figs.~\ref{Fig1_FS}(a)-\ref{Fig1_FS}(b).

For crystals with the D$_{4h}$ point group symmetry, the XX, X$^{\prime}$Y$^{\prime}$ and XY Raman geometries probe the A$_{1g}$+B$_{1g}$, A$_{2g}$ + B$_{1g}$ and A$_{2g}$+ B$_{2g}$ channels, respectively \cite{Devereaux2007RMP}. Assuming the same featureless luminescence background $I_{BG}$ for all polarization geometries and that the A$_{2g}$ response is negligible, the imaginary part of the Raman susceptibility in the A$_{1g}$ channel can be obtained by subtracting the X$^{\prime}$Y$^{\prime}$ spectrum from the XX spectrum and then dividing by the Bose-Einstein factor $1+n(\omega,T)$. The imaginary part of the Raman susceptibility in the B$_{1g}$  and B$_{2g}$ channels can be obtained from X$^{\prime}$Y$^{\prime}$ and XY spectra, respectively.

\section{Results}\label{Results}
\subsection{Normal state}\label{Normal state}

\begin{figure*}[!t]
\begin{center}
\includegraphics[width=5in]{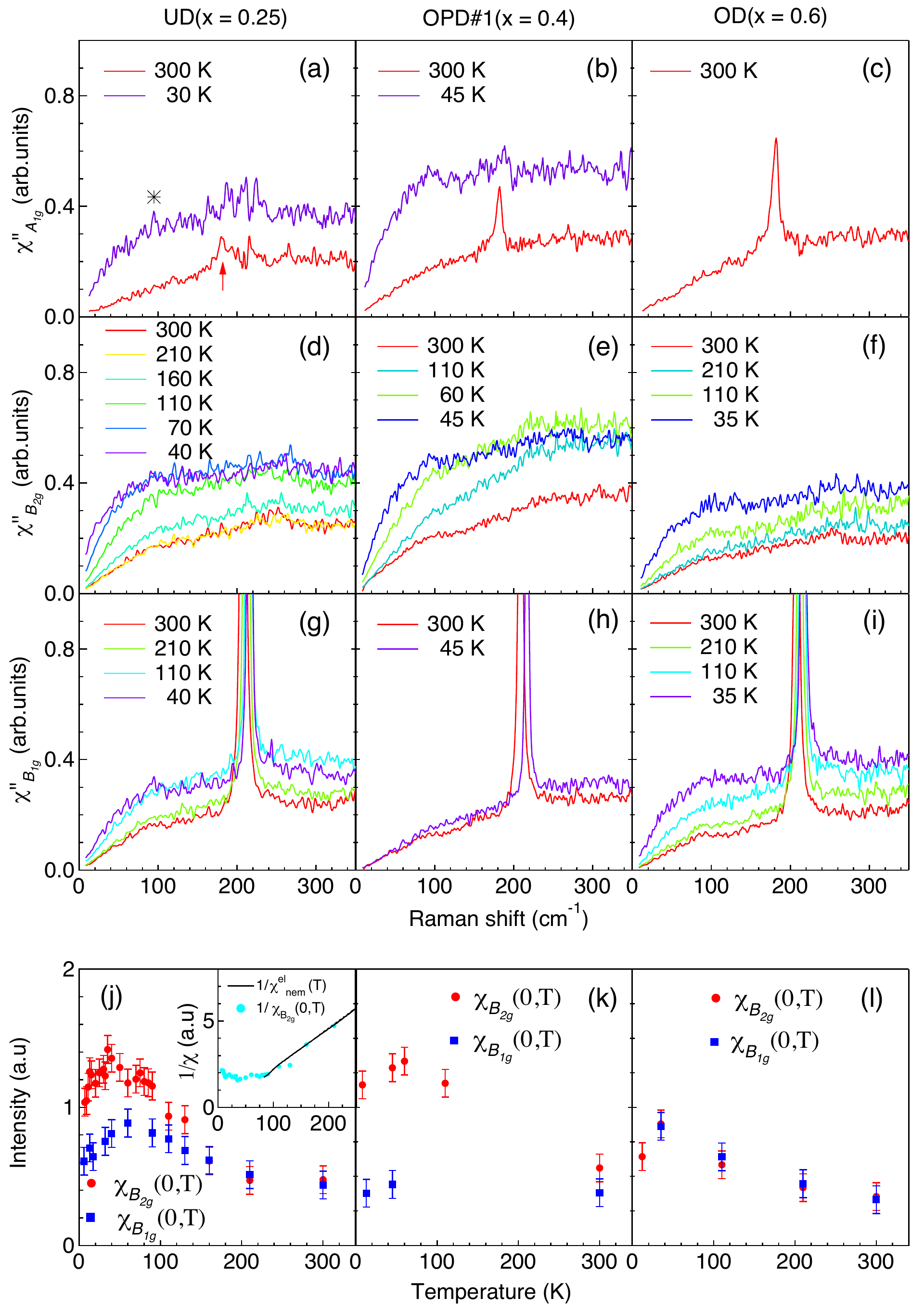}
\end{center}
\caption{\label{Fig2_QEP}(Color online). Doping and temperature evolution of the Raman susceptibility of Ba$_{1-x}$K$_{x}$Fe$_2$As$_2$ in different symmetry channels. Left column:  UD ($x=0.25$); Central column: OPD\#1($x=0.4$); Right column: OD ($x=0.6$). (a)-(c) Temperature dependence of the Raman response in the A$_{1g}$ channel. The asterix in (a) marks a small peak due to laser plasma, whereas the arrow indicates a A$_{1g}$ phonon. (d)-(f) Temperature dependence of the Raman response in the B$_{2g}$ channel. (g)-(i) Same as (d)-(f) but for the B$_{1g}$ channel. (j)-(l) $T$-dependence of the static Raman susceptibilities $\chi_{B_{2g}}(0,T)$ (red solid circles) and $\chi_{B_{1g}}(0,T)$ (blue solid squares).
The inset of (j) shows the inverse nematic susceptibility $1/\chi^{el}_{nem}$ (black line) of Ba$_{0.86}$K$_{0.24}$Fe$_2$As$_2$ extracted from Young's modulus measurements in Ref.\,\cite{Bohmer2014PRL}. The cyan dots in the inset of (j) are values of $1/\chi_{B_{2g}}(0,T)$ derived from Raman.}
\end{figure*}
    
In Figs.~\ref{Fig2_QEP}(a)-\ref{Fig2_QEP}(i), we show the normal state Raman spectra of Ba$_{1-x}$K$_{x}$Fe$_2$As$_2$ in three different symmetry channels. The sharp mode around 182\,~\,cm$^{-1}$ detected at room temperature in Figs.~\ref{Fig2_QEP}(a)-\ref{Fig2_QEP}(c) corresponds to a A$_{1g}$ phonon. The phonon frequency hardens upon cooling\,\,\cite{RahlenbeckPRB80}.  The phonon intensity strengthens with K doping. 
The B$_{2g}$ symmetry electronic continuum strengthens upon cooling from 300\,K to 40\,K [Figs.~\ref{Fig2_QEP}(d)-\ref{Fig2_QEP}(f)].
In particular, at low temperature a broad low-energy feature centered around 100 cm$^{-1}$ develops. Similar quasi-elastic scattering was previously related to quadrupolar nematic fluctuations\,\cite{Thorsmolle2016PRB,ZhangWL2014arxiv}. We  note  that   the  intensity  of this  quasi-elastic  scattering  for  Ba$_{1-x}$K$_{x}$Fe$_2$As$_2$  is weaker than  for Ba(Fe$_{1-x}$Co$_x$)$_2$As$_2$\,\cite{Gallais2013PRL,Kretzschmar2016NatPhy,bohm2016PSSB}, which is possibly due to the different anisotropic  properties  of the  electron-doped and hole-doped Fe-based  superconductors also noted by resistivity  measurements \,\cite{Ying2011PRL,blomberg2013NatCom}. In addition to the B$_{1g}$ phonon at 208\,cm$^{-1}$, the spectra in the B$_{1g}$ symmetry channel also contains quasi-elastic scattering  features similar to the one discussed above  [Figs.~\ref{Fig2_QEP}(g)-\ref{Fig2_QEP}(i)].

In Figs.~\ref{Fig2_QEP}(j)-\ref{Fig2_QEP}(l), we show the static Raman susceptibilities $\chi_{B_{1g}}(0,T)$ and $\chi_{B_{2g}}(0,T)$ obtained \textit{via} the Kramers-Kronig transformation with a high-energy cut-off at 350\,cm$^{-1}$ justified by an already small $\chi''(\omega)/\omega$ integrand at that energy. We used a linear  extrapolation for the $\chi''(\omega)$ below 10 cm$^{-1}$. The B$_{1g}$ phonon was removed by fitting before the Kramers-Kronig transformation. 
The susceptibilities show general enhancement upon cooling from room temperature followed by a mild reduction at low temperatures.
$\chi_{B_{2g}}(0,T)$ is larger than $\chi_{B_{1g}}(0,T)$ in the under-doped [Fig.~\ref{Fig2_QEP} (k)] and optimally-doped [Fig.~\ref{Fig2_QEP} (l)] samples, suggesting that the B$_{2g}$ channel is the dominant channel for the nematic fluctuations.
However, the B$_{1g}$ and B$_{2g}$ symmetry susceptibilities are quite similar in the over-doped regime. 
In a recent study of BaFe$_2$(As$_{0.5}$P$_{0.5}$)$_2$, it was argued that the similarity between the $\chi_{B_{1g}}(0,T)$ and $\chi_{B_{2g}}(0,T)$ static susceptibilities could originate from a disorder due to As/P substitution \cite{wu2016arxiv}. The same argument could also apply here due to the Ba/K substitution.

In the inset of Fig.~\ref{Fig2_QEP}(j), we show the inverse of the static susceptibility $1/\chi_{B_{2g}}(0,T)$ and compare it to the measurements of the elastic modulus  $C_{66}(T)$ \,\cite{GotoJPSJ}. Following the model proposed in the Ref. \cite{Bohmer2014PRL}, $C_{66}(T)$ is renormalized due to the electron-lattice coupling following $C_{66}(T)=C_{66,0}$-$\lambda^2\chi_\phi(T)$, where $C_{66,0}$ is the bare elastic constant, $\phi$ is the nematic order parameter, $\chi_\phi$ is the related nematic susceptibility, $\lambda$ is the electron-lattice coupling constant \cite{Fernandes2010PRL,Bohmer2014PRL}. The electronic nematic susceptibility $\chi^{el}_{nem}(T)$ can thus be derived  from measurements of the elastic modulus $C_{66}(T)$ \cite{GotoJPSJ} (or Young's modulus $Y_{110}(T)$ with $C_{66}/C_{66,0}\approx Y_{110}/Y_{0}$ \cite{Bohmer2014PRL}.)
As shown in the inset of Fig.~\ref{Fig2_QEP}(j), $1/\chi_{B_{2g}}(0,T)$ from Raman measurements scales satisfactorily with $1/\chi^{el}(T)$ computed and scaled from Young's modulus measurements of Ba$_{0.86}$K$_{0.24}$Fe$_2$As$_2$ from Ref.\,\cite{Bohmer2014PRL}.
This scaling above $T_S$ in the under-doped regime suggests that the softening of $C_{66}$ \cite{GotoJPSJ} and the enhancement of the Raman static susceptibility upon cooling are related.

\subsection{Superconducting state}\label{Superconducting state}

\begin{table}[!t]
\caption{\label{gap} Summary of the SC gaps and bosonic modes deduced from Raman scattering, ARPES, STS and inelastic neutron scattering (INS). UD, OPD and OD refer to under-doped, optimally-doped and over-doped samples, respectively. We caution that the doping of the under-doped and over-doped samples measured by different techniques may be different and that the collective modes observed by Raman and by other types of spectroscopies may have different origins. All energies  are given in units of meV.}
\begin{ruledtabular}
\begin{tabular}{cccccc}
&Raman&Raman&ARPES&STS&INS\\
&(This work)&(\cite{Rudi_2013PRL,Rudi_2014PRX})&&&\\
\hline
$\Delta_\alpha^{\textrm{(UD)}}$&&& 9\,\cite{xu2011NatCom}&6\,\cite{Jing2015CPB}&\\
$\Delta_\beta^{\textrm{(UD)}}$&3.8&&4\,\cite{xu2011NatCom}&3.8\,\cite{Jing2015CPB}&\\
$E_{CM}^{\textrm{(UD)}}$&12&&&8\,\cite{Jing2015CPB}&12.5\,\cite{Castellan2011PRL}\\
$\Delta_\alpha^{\textrm{(OPD)}}$&10.8 (B$_{2g}$)&10.6&9-13\,\cite{ding2008EPLgap,L_Zhao,nakayama2009EPLgap}&10.5\,\cite{YinJX2016arxiv}&\\
$\Delta_\beta^{\textrm{(OPD)}}$&4.4&4.4&5-6\,\cite{ding2008EPLgap,L_Zhao,nakayama2009EPLgap}&6\,\cite{YinJX2016arxiv}&\\
$E_{CM}^{\textrm{(OPD)}}$&17.5&17.5&13$\pm$2\,\cite{Richard2009PRLkink}&14\,\cite{shan2012PRLSTM}&14\,\cite{christianson2008NatureINS}\\
$\Delta_\alpha^{\textrm{(OD)}}$&10&& 8\,\cite{Nakayama_PRB2011}&6\,\cite{YinJX2016arxiv}&\\
$\Delta_\beta^{\textrm{(OD)}}$&3&&4\,\cite{Nakayama_PRB2011}&3\,\cite{YinJX2016arxiv}&\\
$E_{CM}^{\textrm{(OD)}}$&14&&&&12\,\cite{Lee2016SciRep}\\
\end{tabular}
\end{ruledtabular}
\begin{raggedright}
\end{raggedright}
\end{table}

Before discussing the Raman scattering features observed at low temperature, we recall the SC gap values obtained by complementary spectroscopic probes in optimally-doped Ba$_{1-x}$K$_{x}$Fe$_2$As$_2$. ARPES studies report nodeless SC gaps on all Fermi surface (FS) pockets, with small or negligible in-plane anisotropy\,\cite{ding2008EPLgap,L_Zhao}. While a SC gap of 6 meV is found on the holelike $\beta$ ($d_{xy}$) FS centered at the $\Gamma$ point, a larger gap of about 12 meV is found on all the other pockets, with differences smaller than a meV\,\cite{nakayama2009EPLgap}. An ARPES study of the SC gap using synchrotron radiation, which allows to vary the $k_z$ position, indicates that the gap size on each FS does not vary significantly with $k_z$, except for the $\Gamma$-centered hole FS formed by the even combination of the $d_{xz}$ and $d_{yz}$ orbitals, for which a gap varies between 9 and 12 meV\,\cite{xu2011NatPhysics}. Results compatible with ARPES are obtained by STS, which reveals two coherence SC peaks at 10.5~meV and 6~meV\,\cite{YinJX2016arxiv}, and by optical conductivity, for which a SC gap of 12.5~meV opens below $T_c$\,\cite{Li2008PRL}. Thermal conductivity measurements are consistent with nodeless gaps for the optimally-doped compound \cite{XG_LuoPRB2009}. At the energy scale similar to the SC gaps, a 14~meV neutron resonance mode is reported below $T_c$ at the antiferromagnetic wave-vector $\mathbf{Q}$\,\cite{christianson2008NatureINS}. Interestingly, a 13$\pm$2 meV mode energy determined from a kink in the electronic dispersion is observed by ARPES below $T_c$ on bands quasi-nested by the antiferromagnetic wave-vector\,\cite{Richard2009PRLkink}. STS measurements also reveal coupling to a bosonic mode at 14 meV\,\cite{shan2012PRLSTM}. We summarize values of the SC gaps and bosonic modes deduced from different spectroscopies in TABLE \ref{gap}.

\subsubsection{The optimal doping}

\begin{figure}[!t]
\begin{center}
\includegraphics[width=3.4in]{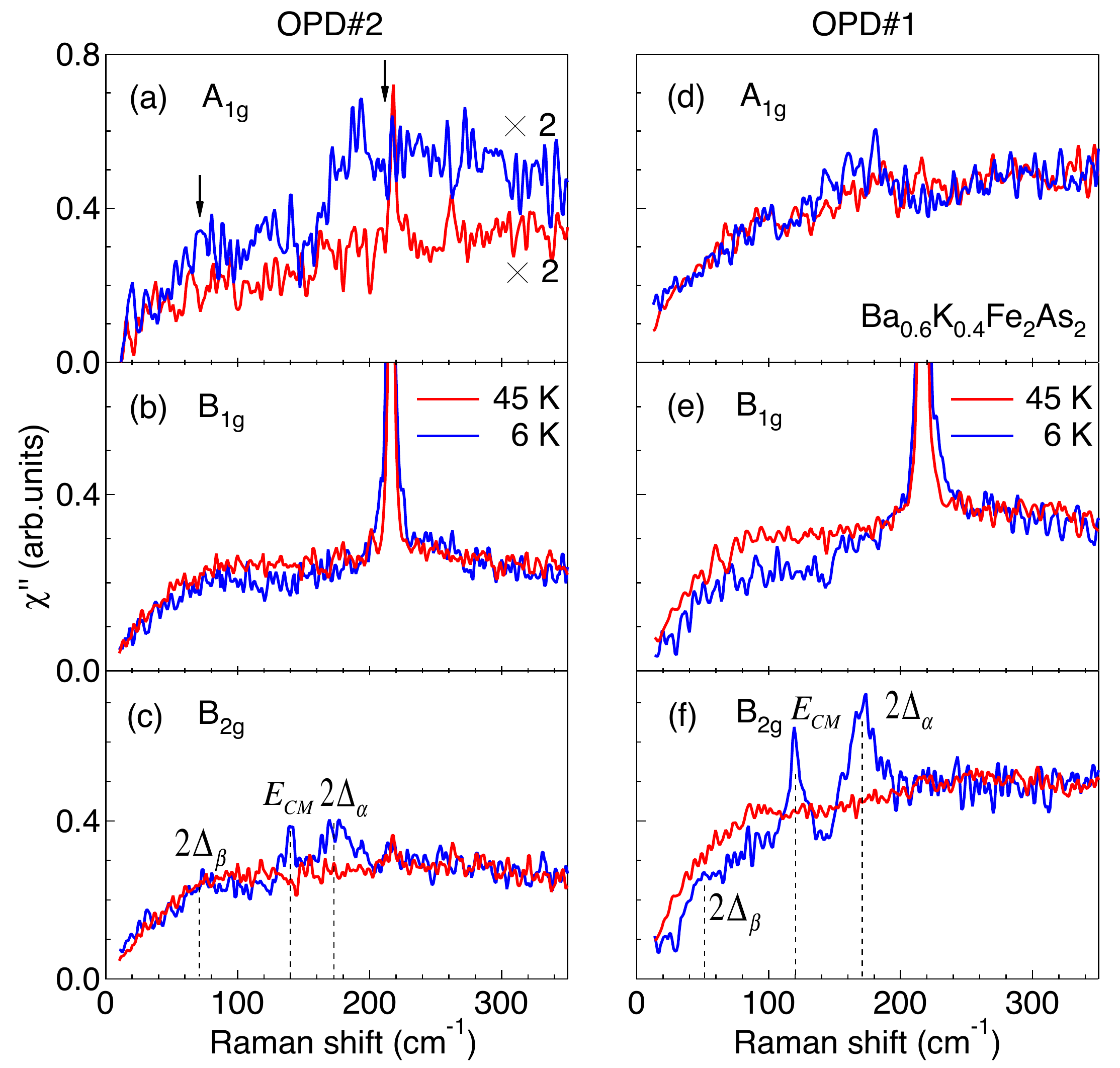}
\end{center}
\caption{\label{Fig3_OPD}(Color online) (a)-(c) Raman response of Ba$_{0.4}$K$_{0.6}$Fe$_2$As$_2$ (OPD\#2) at 45\,K (red) and 6\,K (blue) in the (a) A$_{1g}$, (b) B$_{1g}$, and (c) B$_{2g}$ channels. The dashed lines in (c) mark a broad feature at 70\,cm$^{-1}$ ($2\Delta_{\beta}$), a collective mode ($E_{CM}$) around 140\,cm$^{-1}$ and a pair-breaking peak at 172\,cm$^{-1}$ ($2\Delta_{\alpha}$). (d)-(f) Same as (a)-(c) but for sample OPD\#1. For the OPD\#1 sample we find $2\Delta_{\beta}=50$~cm$^{-1}$, $E_{CM}=120$~cm$^{-1}$ and $2\Delta_{\alpha}=168$~cm$^{-1}$. }
\end{figure} 
    
In Fig.~\ref{Fig3_OPD}, we compare the Raman spectra at 45\,K (normal state) and 6\,K (SC state) in three symmetry channels from optimally-doped samples OPD\#1 and OPD\#2. We first start describing results from the OPD\#2 sample. In Fig.~\ref{Fig3_OPD}(a), two broad and weak features emerge around 70\,cm$^{-1}$ and 210\,cm$^{-1}$, which we assign to A$_{1g}$ SC pair breaking peaks corresponding to gap values 2$\Delta$ of 8.8~meV and 26.2~meV, respectively. In Fig.~\ref{Fig3_OPD}(b), a small spectral weight suppression is seen below 160\,cm$^{-1}$ in the B$_{1g}$ channel. In Fig.~\ref{Fig3_OPD}(c), a broad and weak feature at 70\,cm$^{-1}$ (8.8~meV) is observed in the B$_{2g}$ channel, which we assign to the small gap 2$\Delta_\beta$ on the $\beta$ FS pocket with $d_{xy}$ character\,\cite{ding2008EPLgap,L_Zhao}. Another sharp mode at 172\,cm$^{-1}$ associated with a SC pair breaking peak at 2$\Delta_\alpha=21.6$~meV appears in the B$_{2g}$ channel, which is consistent with the 10-13~meV magnitude measured by ARPES for the large SC gap around $k_z=0$\,\cite{ding2008EPLgap,L_Zhao,nakayama2009EPLgap}. The large gap value varies from 10.8~meV in the B$_{2g}$ channel to 13.1\,meV in the A$_{1g}$ channel, in agreement with ARPES measurements revealing an anisotropic gap along $k_z$\,\cite{xu2011NatPhysics}. Between 2$\Delta_\beta$ and 2$\Delta_\alpha$, we detect a sharp mode at $E_{CM}=140$\,cm$^{-1}$ (17.5~meV), which will be discussed below. 
   
For the OPD\#1 sample, only a small broad feature around  160\,cm$^{-1}$ is seen in the A$_{1g}$ symmetry response [Fig. 3(d)].  In contrast, the spectral features in the B$_{1g}$ and B$_{2g}$ channels appear more clearly for the OPD\#1 sample than for the OPD\#2 sample. In Fig.~\ref{Fig3_OPD}(e), a spectral weight suppression below $T_c$ is seen below 160\,cm$^{-1}$ in the B$_{1g}$ channel. For the B$_{2g}$ channel, two sharp modes at 120\,cm$^{-1}$ and 168\,cm$^{-1}$, as well as a kink feature at 50\,cm$^{-1}$, are seen in Fig.~\ref{Fig3_OPD}(f). While little change is observed for the large SC gap pair breaking peak energy as compared to the OPD\#2 sample, a substantial shift from 70 cm$^{-1}$ to 50\,cm$^{-1}$ is observed for the small SC gap pair breaking peak energy. The sharp $E_{CM}$ mode shifts by the same amount, from 140\,cm$^{-1}$ to 120\,cm$^{-1}$ in the OPD\#1 sample. Since the results for the OPD\#2 sample are consistent with previous Raman work\,\cite{Rudi_2013PRL} for the optimally-doped compound, we caution that the OPD\#1 sample cleaved in this study must have a slightly different doping due to inhomogeneous K distribution in the bulk or rapid sample aging.

In addition to the sharp peak, a threshold is also observed around 30\,cm$^{-1}$ in the SC state [Figs.~\ref{Fig3_OPD}(e) and \ref{Fig3_OPD}(f)]. This threshold suggests a fundamental gap of 1.9~meV, consistent with the 2 meV-wide flat bottom in the STS spectra\,\cite{YinJX2016arxiv}. No clear threshold is detected in the OPD\#2 sample though, possibly because the cleaved surface is not good enough, as suggested by weaker peaks in the B$_{2g}$ channel.

\subsubsection{The over-doped regime}  
\begin{figure}[!t]
\begin{center}
\includegraphics[width=2.6in]{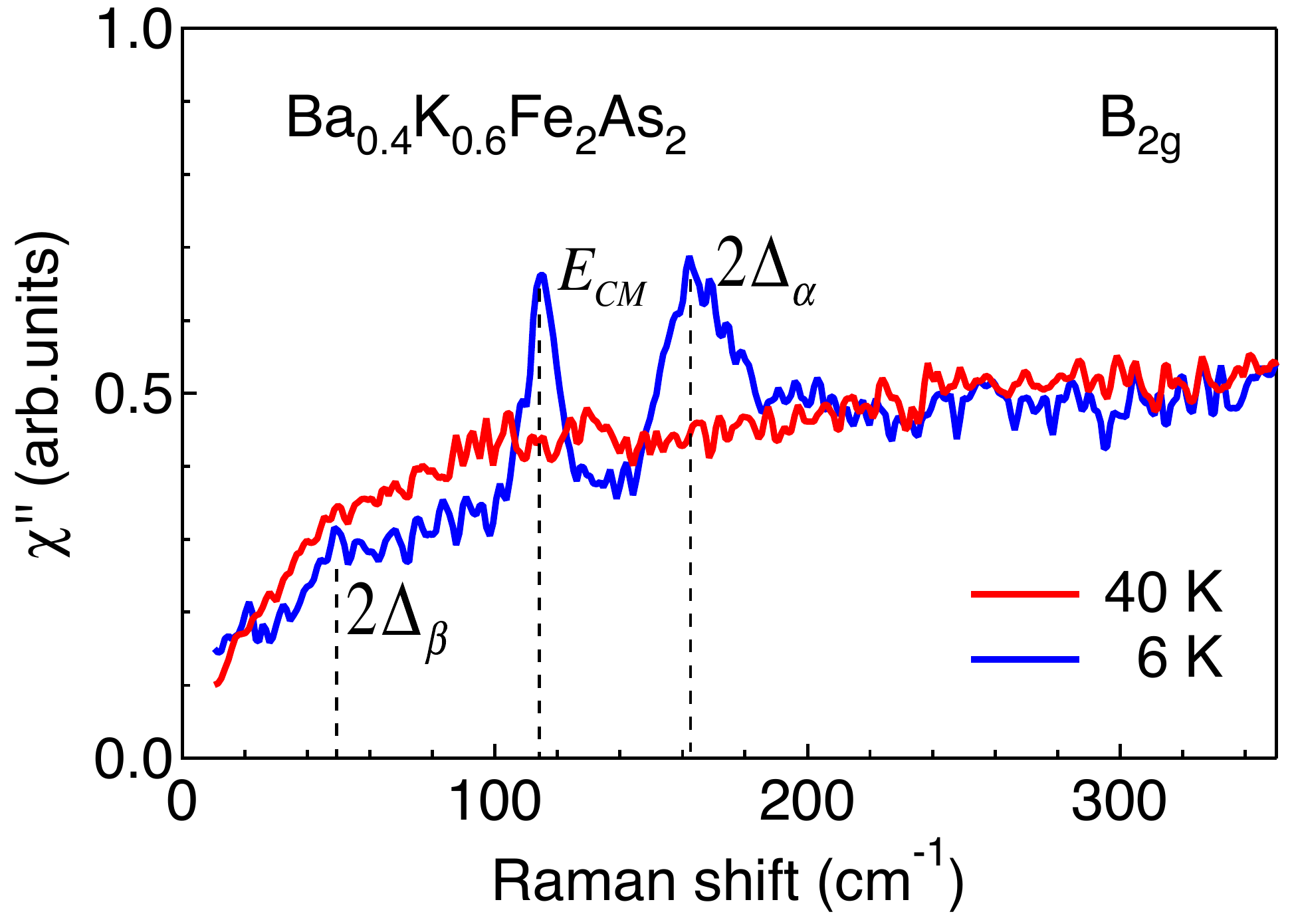}
\end{center}
\caption{\label{Fig4_OD}(Color online) Raman response of Ba$_{0.6}$K$_{0.4}$Fe$_2$As$_2$ (OD) at 40\,K (red) and 6\,K (blue) in the B$_{2g}$ channel. The dashed lines mark a broad peak at 50\,cm$^{-1}$ ($2\Delta_{\beta}$), a collective mode $E_{CM}$ at 115\,cm$^{-1}$ and a pair breaking peak at 162\,cm$^{-1}$ ($2\Delta_{\alpha}$).}
\end{figure}

We now discuss the spectra from  the over-doped sample. In Fig.~\ref{Fig4_OD}, we compare the Raman response obtained at 40\,K (normal state) and 6\,K (SC state) in the B$_{2g}$ channel. Four features are clearly observed: a threshold around 30\,cm$^{-1}$, a kink-like feature around 50\,cm$^{-1}$, and two sharp modes at 115\,cm$^{-1}$ and 162\,cm$^{-1}$. As with the OPD\#1 sample, we assign the threshold to a fundamental SC gap. The kink around 50\,cm$^{-1}$ corresponds to the small SC gap pair breaking peak with 2$\Delta_\beta=6$ meV. The sharp mode at 162\,cm$^{-1}$ corresponds to the large SC gap pair breaking peak with 2$\Delta_\alpha=20$ meV. As a comparison, an ARPES study on over-doped Ba$_{0.7}$K$_{0.3}$Fe$_2$As$_2$ ($T_c=22$~K) gives $\Delta_\alpha=8$~meV and $\Delta_\beta=4$~meV \cite{Nakayama_PRB2011}. Finally, the sharp mode at 115\,cm$^{-1}$ (14~meV) is associated to the $E_{CM}$ mode. We note that all the features for the OD sample are similar to those for the OPD\#1 sample, confirming that the OPD\#1 sample might be slightly over-doped.
 
\subsubsection{The under-doped regime}

\begin{figure}[!t]
\begin{center}
\includegraphics[width=3.4in]{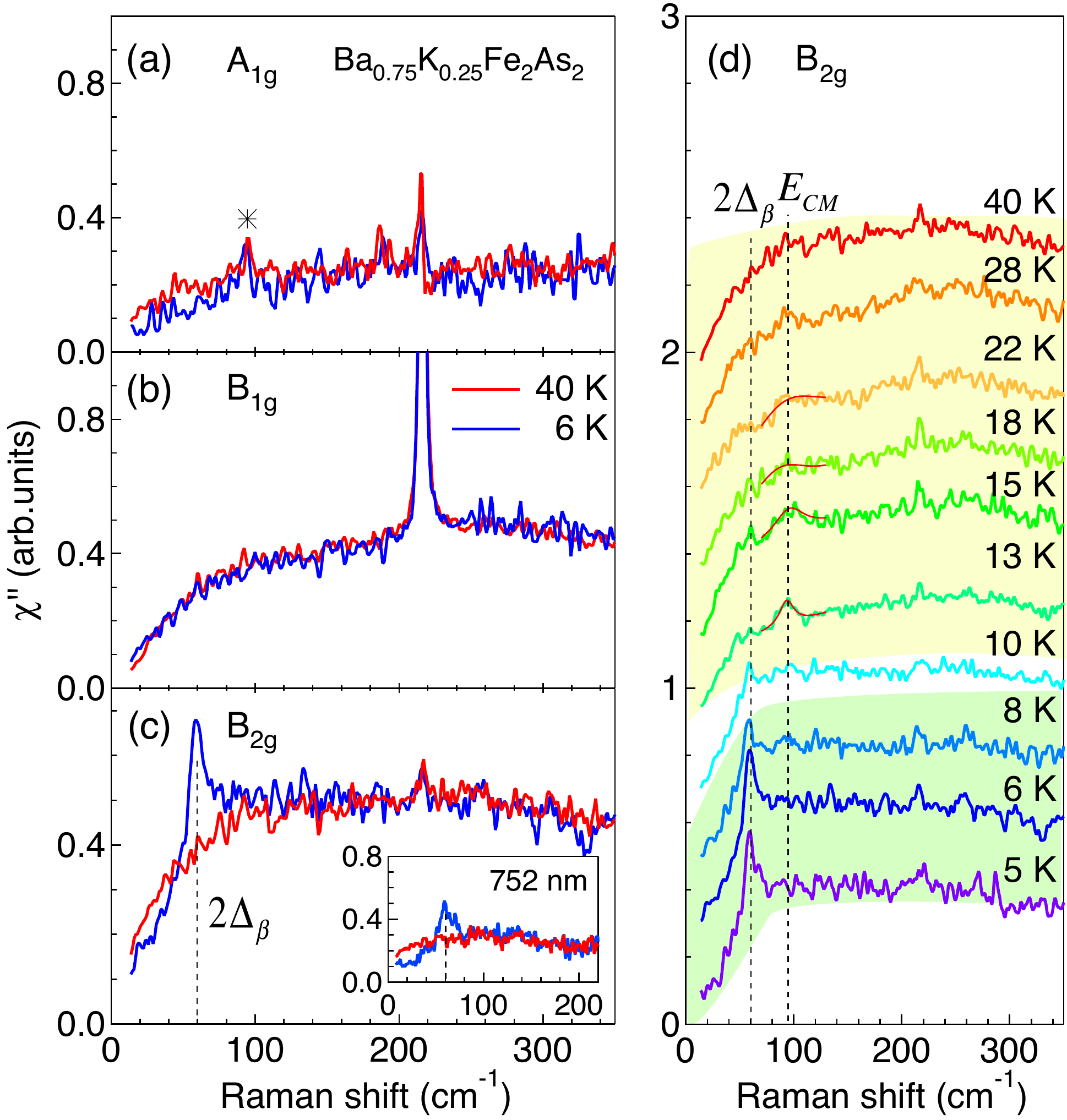}
\end{center}
\caption{\label{Fig5_UD}(Color online). Raman response of Ba$_{0.75}$K$_{0.25}$Fe$_2$As$_2$ (UD30K) at 40\,K (red) and 6\,K (blue)  for the (a) A$_{1g}$, (b) B$_{1g}$, and (c) B$_{2g}$ symmetries. The star in (a) represents a laser plasma line. The inset in (c) shows the Raman responses recorded with a 752\,nm laser excitation. (d) $\chi''_{B_{2g}}(\omega)$ at various temperatures. The dashed lines in (d) indicate $2\Delta_{\beta}$ and $E_{CM}$. The red curves in (d) are fits of the $E_{CM}$ peaks. The yellow and green shadings emphasize different spectral backgrounds associated to different phases.}
\end{figure}

We now discuss results for the under-doped regime. In the left column of Fig.~\ref{Fig5_UD}, we compare the Raman responses $\chi''(\omega)$ from the under-doped sample at 40\,K (normal state) and 6\,K (SC state) in three symmetry channels.  A small suppression of spectral weight is observed below $T_c$ at low energies in the A$_{1g}$ channel [Fig.~\ref{Fig5_UD}(a)], and the spectra barely change in the B$_{1g}$ channel [Fig.~\ref{Fig5_UD}(b)]. In the B$_{2g}$ channel, however, spectral weight is transferred from the low-energy, and a sharp peak at 60\,cm$^{-1}$ builds up. This peak is also seen when 752\,nm excitation is used, as shown in the inset of Fig.~\ref{Fig5_UD}(c). Following the interpretation of the kink observed at 70\,cm$^{-1}$ at optimal doping, we attribute the 60\,cm$^{-1}$ feature in the UD sample to a pair breaking peak with 2$\Delta_\beta=7.5$ meV, which is consistent with the $\Delta_\beta=4$ meV gap value reported by ARPES measurements for the $\beta$ ($d_{xy}$) $\Gamma$-centered hole FS pocket for samples with similar doping level\,\cite{xu2011NatCom}. Surprisingly, the sharp SC pair breaking peak at 172\,cm$^{-1}$ observed at low temperature for optimally-doped samples is absent in the UD sample. Although the reason for this disappearance is unclear, we caution that it may be related to the loss of coherence observed by ARPES experiments for the $d_{xz}/d_{yz}$ bands\,\cite{xu2011NatCom}.  

As illustrated by the fine temperature dependence of the B$_{2g}$ Raman response in Fig.~\ref{Fig5_UD}(d), the sharp peak at 60\,cm$^{-1}$ appears clearly only below 10\,K. Interestingly, the B$_{2g}$ spectrum exhibits clear changes across that temperature, as highlighted with yellow and green backgrounds in Fig.~\ref{Fig5_UD}(d). For example, below 10\,K the spectral background is flat between 100\,~\,cm$^{-1}$ and 350\,~\,cm$^{-1}$, but shows a broad feature above that temperature. These observations are consistent with recent studies on Ba$_{1-x}$K$_{x}$Fe$_2$As$_2$\,\cite{bohmer2015NatCom,Allred_PRB92} and Ba$_{1-x}$Na$_{x}$Fe$_2$As$_2$ \cite{Khalyavin_PRB90,L_Wang_PRB93} suggesting re-entrance into the $C_4$ preserved magnetic phase in the under-doped regime. Within this context, the broad feature above 10~K can be interpreted as the formation of a spin-density-wave gap below the magnetic phase transition. We note that a pseudo-gap of about 17 meV was observed by ARPES below 125~K in under-doped Ba$_{0.75}$K$_{0.25}$Fe$_2$As$_2$ \cite{xu2011NatCom}. Assuming that this pseudo-gap is approximately symmetric with respect to the Fermi energy, it would lead to a Raman feature at twice this value ($\sim$ 35 meV), which is roughly the position of the broad feature observed in Raman data. The sudden disappearance of the broad feature below 10~K could be explained either by a non-magnetic low-temperature phase ($T<10$~K), which would contradict the phase diagram presented in Ref. \cite{bohmer2015NatCom}, by a different magnetic structure, or by restoring the four-fold symmetry at the lowest temperature. The $E_{CM}$ mode in the UD sample is detected around 95\,~\,cm$^{-1}$ only between 22\,K and 13\,K, emphasizing further the difference between the phases above and below the phase transition at 10 K. 
  
\section{Discussion}\label{Discussion}
   
\begin{figure}[!t]
\begin{center}
\includegraphics[width=3.4in]{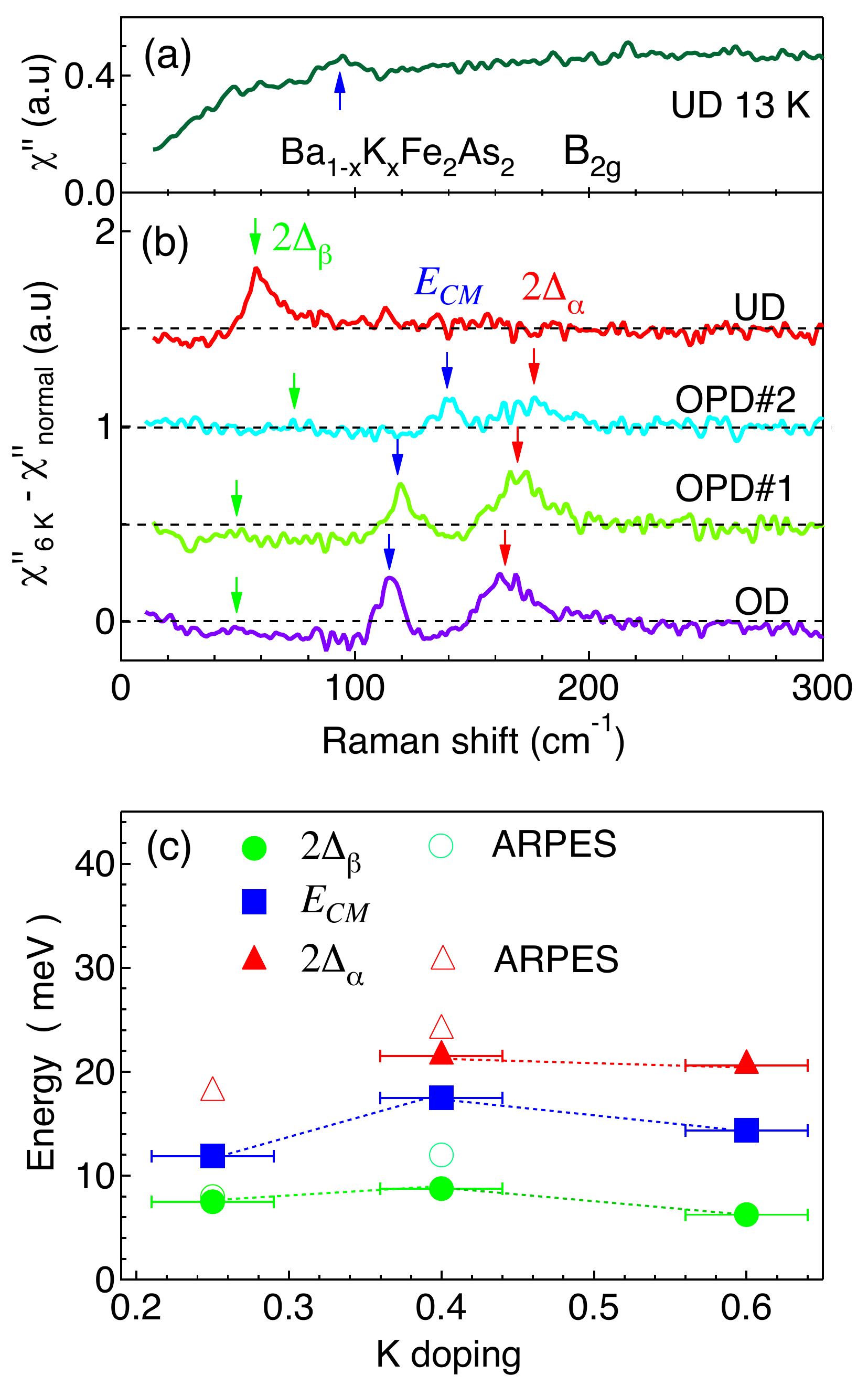}
\end{center}
\caption{\label{Fig6_summary}(Color online). (a) Raman response of Ba$_{0.75}$K$_{0.25}$Fe$_2$As$_2$ in the B$_{2g}$ channel at 13\,K. (b) Difference between the Raman spectra at 6\,K in the SC state and in the normal state, recorded in the B$_{2g}$ channel for different dopings. (c) Summary of the SC pair breaking peaks and in-gap mode in Ba$_{1-x}$K$_{x}$Fe$_2$As$_2$ obtained in the B$_{2g}$ channel. The full and open symbols correspond to results from this work and from ARPES\,\cite{ding2008EPLgap,L_Zhao,xu2011NatCom}, respectively.}
\end{figure}
   
In this section we discuss the origin of the $E_{CM}$ mode. In Fig.~\ref{Fig6_summary}(b), we plot the doping dependence of the difference between the B$_{2g}$ Raman response function recorded in the SC state at 6 K, deep in the SC state, and at the normal state. Although both the  $2\Delta_{\alpha}$ and $2\Delta_{\beta}$ peaks shift with doping, the shift is more pronounced for the later one [see Fig.~\ref{Fig6_summary}(c)]. Interestingly, the $E_{CM}$ mode moves almost by the same amount as the $2\Delta_{\beta}$ peak: the mode is observed at 95\,cm$^{-1}$ (11.9 meV) for $x=0.25$, at 140\,cm$^{-1}$ (17.5 meV) in for $x=0.4$ and at 115\,cm$^{-1}$ (14.4 meV) for $x=0.6$ doping levels.
The $E_{CM}$  mode energy is higher than the gap typically observed by ARPES  for the $\beta$ ($d_{xy}$) band and smaller than the gap observed on the other FSs for corresponding dopings \,\cite{xu2011NatCom,ding2008EPLgap,L_Zhao}. Consequently,  the $E_{CM}$ mode is unlikely related  to a SC pair breaking peak on the same band.

\begin{table}[!t]
\caption{\label{BindingEnergy} Summary of the binding energy of the in-gap mode in the B$_{2g}$ channel for the OPD and OD samples. All values are given in cm$^{-1}$. }
\begin{ruledtabular}
\begin{tabular}{ccccc}
Sample&$2\Delta_\alpha$&$E_{CM}$&$E_{B}$&$E_{B}/2\Delta_\alpha$\\
\hline
OPD\#2&172&140& 32&0.2\\
OD&162&115&47&0.3\\
\end{tabular}
\end{ruledtabular}
\begin{raggedright}
\end{raggedright}
\end{table}

We note that the $E_{CM}$ mode energy is similar to the sum $\Delta_\beta+\Delta_\alpha$. One speculative explanation for the $E_{CM}$ related  to  an  inter-band scattering  process lies in the  observation of in-gap impurity states  by ARPES  below $T_c$\,\cite{Zhang2014PRXimpurity}: a photon  breaks  a Cooper  pair out of the condensate and creates a quasi-particle on the  band with an energy cost $\Delta_\alpha$, while the second particle from the broken pair is scattered into a quasi-particle state of the band with the smaller gap (energy cost $\Delta_\beta$), with the help of an impurity taking the recoil for conservation of the quasi-momentum. Due to the residual interaction coming from both pairing and Coulomb interaction between two quasi-particles on different bands, and to some charge transfer between bands, the cost of this process is slightly smaller than $\Delta_\alpha$+$\Delta_\beta$. However, it is not clear within this scenario why the related Raman mode is so sharp and symmetric.

We note that the energy of the $E_{CM}$ mode is similar to that of the neutron resonance mode observed only below $T_c$ in the triplet channel at the antiferromagnetic wave vector\,\cite{christianson2008NatureINS} and to the kink energy observed in the quasiparticle dispersion by ARPES also only below $T_c$\,\cite{Richard2009PRLkink}. In principle, only spin singlet modes with nearly zero momentum transfer can be probed by Raman scattering. Thus, the neutron resonance mode and the Raman collective mode are distinct. The fact that the binding energies of these two modes are similar suggests that the interaction leading to the origin of in-gap resonance in the magnetic channel is not that different from the attraction in the spin singlet channel. In other words, interactions at momentum transfer $\mathbf{q}=0$ and $\mathbf{q}=\mathbf{Q}$ (such as intra-pocket and inter-pocket interactions, respectively), have similar strength. Hence, a proper model description of the collective modes in such superconductor, must consider both types of interactions on equal footing.
 
In Table \ref{BindingEnergy}, we summarize the binding energy $E_B=2\Delta_{\alpha}-E_{CM}$ and the ratio between  the binding energy and the large gap edge $E_{B}/2\Delta_\alpha$ for OPD\#2 and OD samples. With doping  the binding energy increases from 32 cm$^{-1}$  for optimally-doped  regime to 47 cm$^{-1}$ for the over-doped regime, and the ratio $E_{B}/2\Delta_\alpha$  increases from 0.2 to 0.3, indicating enhancement of the residual interactions with doping. 

The interaction could originate from the attraction in sub-dominant symmetry particle-particle channel leading to a Bardasis-Schrieffer (BS) like exciton \cite{BS1961PR,Klein1984,Maiti2015PRB,Maiti2016arxiv,Thorsmolle2016PRB,Khodas2014PRB} or, alternatively, from particle-hole attraction leading to nematic fluctuations and a Pomeranchuk-like exciton \cite{Klein_PRB82,Thorsmolle2016PRB, Gallais2016PRL, Khodas2014PRB,Chubukov2016PRB}.  The increase of the binding energy with doping within the first BS scenario is an indication that the competing $d$-wave symmetry interaction strengthen with doping. Indeed, although fully gapped superconductivity is well established in the optimally doped regime, numerous experiments suggest that transition from nodeless to nodal order parameter appear in the heavily hole-doped regime for $x>0.8$~\cite{Shibauchi2014QCP,Fukazawa2009JPSJ,Hashimoto2010PRB,Reid2012PRL,Okazaki1314}. Because the structural instability is suppressed with K-doping, the nematic interactions weaken, in agreement with the observed reduction of the nematic susceptibility with doping [Fig.~\ref{Fig2_QEP} (j-l)].

However, the  nematic  fluctuations  can grow stronger below $T_c$,  where  low-lying excitations  are  gapped  and  thus  the  damping  of the  nematic fluctuations  is removed.  In this case nematic  fluctuations  can gain coherence and lead to a particle-hole exciton mode manifesting itself as a sharp resonance in the B$_{2g}$  channel~\cite{Thorsmolle2016PRB, Gallais2016PRL,Chubukov2016PRB}. Interestingly, the  collective modes that appear  in the tetragonal phase of the optimally-doped  and over-doped samples are sharper and stronger than  that  in the orthorhombic  phase of the under-doped  regime, likely due to suppressed  nematic  fluctuations  in the orthorhombic  phase,  where the four-fold symmetry is broken.

We note that for a multi-band system the interactions of the particle-particle and particle-hole channels of the same symmetry representation mix. Therefore, the separation between Bardasis-Schrieffer-like  and Pomeranchuk-like excitons is artificial as both the particle-particle and particle-hole interactions contribute to formation of the in-gap exciton~\cite{Khodas2014PRB}. We also note that  recent theoretical  studies show that  nematic  fluctuations  can enhance  the  $s$-wave Cooper pairing and thus explain the enhancement of $T_c$ near the nematic quantum critical point \cite{Lederer2015PRL,Kang2016PRL}.

\section{Conclusions}\label{Conclusions}

In conclusion, we used polarization-resolved electronic Raman spectroscopy to probe the electronic properties of Ba$_{1-x}$K$_{x}$Fe$_2$As$_2$ in the normal and SC states as a function of doping (0.25$\leq$x$\leq$0.6). We find that temperature dependent quadrupolar nematic  fluctuations are universally present for all studied doping range. The derived dynamic response static Raman susceptibility $\chi_{B_{2g}}(0,T)$ is larger than $\chi_{B_{1g}}(0,T)$, suggesting that nematic fluctuations of the XY symmetry dominate. In particular, the temperature dependence of the static Raman susceptibility $\chi_{B_{2g}}(0,T)$ in the under-doped sample is consistent with measurements of the elastic modulus $C_{66}(T)$, suggesting that the $XY$-symmetry electronic fluctuations and the lattice are strongly coupled.
  
In the SC state, for the optimally doped regime, we detected three features in the B$_{2g}$ symmetry Raman response: two pair breaking peaks at 70~cm$^{-1}$ ($8.75$~meV) and 172~cm$^{-1}$ (21.5~meV) corresponding to a small and a large gap, and an in-gap collective mode at 140~cm$^{-1}$ (17.5 meV). For the over-doped regime, similar three features in B$_{2g}$ channel were observed: two pair breaking peaks at 50~cm$^{-1}$ ($6.25$~meV) and 115~cm$^{-1}$(14.38~meV), and an in-gap mode at 162~cm$^{-1}$ (20.25~meV). We discuss scenarios  for the origin of the in-gap modes including the mixture of Bardasis-Schrieffer-like and Pomeranchuk-like excitons. The binding energy of the in-gap mode increases from optimal doping to over-doping, suggesting a possible transition from nodeless $s_\pm$ order parameter to a nodal $d$-wave order parameter at higher K doping concentration.

In the under-doped regime, the B$_{2g}$ symmetry pair breaking peak corresponding to the large gap is undetectable. We detected a sharp pair breaking peak at 60~cm$^{-1}$ (3.8~meV) corresponding to the small gap. In addition, the shape of the spectral background changes at 10~K, suggesting two distinct SC phases in the under-doped regime. We observed a broader peak at 95~cm$^{-1}$ above 10~K, which we assign to the collective in-gap mode in the under-doped regime.

\section*{Acknowledgments}

We acknowledge useful discussions with K. Haule, V. K. Thorsm{\o}lle, W.-L. Zhang, P. Zhang, H. Miao, and J.-X.Yin. We acknowledge H.-H. Kung and B. Dennis for help with the experiments. The spectroscopic study and analysis  at Rutgers were supported by the US Department of Energy, Basic Energy Sciences, and Division of Materials Sciences and Engineering under Grant No. DE-SC0005463. The materials characterization at IOP, was supported by grants from MOST (2015CB921301, 2016YFA0401000, 2016YFA0300300) and NSFC (11274362, 11674371) of China. The crystal growth was supported by grants from MOST(2011CBA00102) and NSFC(11534005) of China.

\bibliography{biblio_long}

\begin{thebibliography}{70}%
\makeatletter
\providecommand \@ifxundefined [1]{%
 \@ifx{#1\undefined}
}%
\providecommand \@ifnum [1]{%
 \ifnum #1\expandafter \@firstoftwo
 \else \expandafter \@secondoftwo
 \fi
}%
\providecommand \@ifx [1]{%
 \ifx #1\expandafter \@firstoftwo
 \else \expandafter \@secondoftwo
 \fi
}%
\providecommand \natexlab [1]{#1}%
\providecommand \enquote  [1]{``#1''}%
\providecommand \bibnamefont  [1]{#1}%
\providecommand \bibfnamefont [1]{#1}%
\providecommand \citenamefont [1]{#1}%
\providecommand \href@noop [0]{\@secondoftwo}%
\providecommand \href [0]{\begingroup \@sanitize@url \@href}%
\providecommand \@href[1]{\@@startlink{#1}\@@href}%
\providecommand \@@href[1]{\endgroup#1\@@endlink}%
\providecommand \@sanitize@url [0]{\catcode `\\12\catcode `\$12\catcode
  `\&12\catcode `\#12\catcode `\^12\catcode `\_12\catcode `\%12\relax}%
\providecommand \@@startlink[1]{}%
\providecommand \@@endlink[0]{}%
\providecommand \url  [0]{\begingroup\@sanitize@url \@url }%
\providecommand \@url [1]{\endgroup\@href {#1}{\urlprefix }}%
\providecommand \urlprefix  [0]{URL }%
\providecommand \Eprint [0]{\href }%
\providecommand \doibase [0]{http://dx.doi.org/}%
\providecommand \selectlanguage [0]{\@gobble}%
\providecommand \bibinfo  [0]{\@secondoftwo}%
\providecommand \bibfield  [0]{\@secondoftwo}%
\providecommand \translation [1]{[#1]}%
\providecommand \BibitemOpen [0]{}%
\providecommand \bibitemStop [0]{}%
\providecommand \bibitemNoStop [0]{.\EOS\space}%
\providecommand \EOS [0]{\spacefactor3000\relax}%
\providecommand \BibitemShut  [1]{\csname bibitem#1\endcsname}%
\let\auto@bib@innerbib\@empty
\bibitem [{\citenamefont {Mazin}\ and\ \citenamefont
  {Schmalian}(2009)}]{MazinPhysicaC2009}%
  \BibitemOpen
  \bibfield  {author} {\bibinfo {author} {\bibfnamefont {I.~I.}\ \bibnamefont
  {Mazin}}\ and\ \bibinfo {author} {\bibfnamefont {J.}~\bibnamefont
  {Schmalian}},\ }\bibfield  {title} {\enquote {\bibinfo {title} {Pairing
  symmetry and pairing state in ferropnictides: Theoretical overview},}\ }\href
  {http://ac.els-cdn.com/S0921453409001002/1-s2.0-S0921453409001002-main.pdf?_tid=db536eb0-be83-11e6-b98d-00000aacb35f&acdnat=1481338601_e43359850eded8e427dfd1dec7490b5b}
  {\bibfield  {journal} {\bibinfo  {journal} {Physica C}\ }\textbf {\bibinfo
  {volume} {469}},\ \bibinfo {pages} {614} (\bibinfo {year}
  {2009})}\BibitemShut {NoStop}%
\bibitem [{\citenamefont {Graser}\ \emph {et~al.}(2009)\citenamefont {Graser},
  \citenamefont {Maier}, \citenamefont {Hirschfeld},\ and\ \citenamefont
  {Scalapino}}]{Graser_NJP2009}%
  \BibitemOpen
  \bibfield  {author} {\bibinfo {author} {\bibfnamefont {S.}~\bibnamefont
  {Graser}}, \bibinfo {author} {\bibfnamefont {T.~A.}\ \bibnamefont {Maier}},
  \bibinfo {author} {\bibfnamefont {P.~J.}\ \bibnamefont {Hirschfeld}}, \ and\
  \bibinfo {author} {\bibfnamefont {D.~J.}\ \bibnamefont {Scalapino}},\
  }\bibfield  {title} {\enquote {\bibinfo {title} {Near-degeneracy of several
  pairing channels in multiorbital models for the {F}e pnictides},}\ }\href
  {http://stacks.iop.org/1367-2630/11/i=2/a=025016} {\bibfield  {journal}
  {\bibinfo  {journal} {New J. Phys.}\ }\textbf {\bibinfo {volume} {11}},\
  \bibinfo {pages} {025016} (\bibinfo {year} {2009})}\BibitemShut {NoStop}%
\bibitem [{\citenamefont {Fernandes}\ and\ \citenamefont
  {Chubukov}(2017)}]{Fernandes_Chubukov_review}%
  \BibitemOpen
  \bibfield  {author} {\bibinfo {author} {\bibfnamefont {R.~M.}\ \bibnamefont
  {Fernandes}}\ and\ \bibinfo {author} {\bibfnamefont {A.~V.}\ \bibnamefont
  {Chubukov}},\ }\bibfield  {title} {\enquote {\bibinfo {title} {Low-energy
  microscopic models for iron-based superconductors: a review},}\ }\href
  {http://stacks.iop.org/0034-4885/80/i=1/a=014503} {\bibfield  {journal}
  {\bibinfo  {journal} {Rep. Prog. Phys.}\ }\textbf {\bibinfo {volume} {80}},\
  \bibinfo {pages} {014503} (\bibinfo {year} {2017})}\BibitemShut {NoStop}%
\bibitem [{\citenamefont {Rossat-Mignod}\ \emph {et~al.}(1991)\citenamefont
  {Rossat-Mignod}, \citenamefont {Regnault}, \citenamefont {Vettier},
  \citenamefont {Bourges}, \citenamefont {Burlet}, \citenamefont {Bossy},
  \citenamefont {Henry},\ and\ \citenamefont
  {Lapertot}}]{Rossat-Mignod_PhysicaC185}%
  \BibitemOpen
  \bibfield  {author} {\bibinfo {author} {\bibfnamefont {J.}~\bibnamefont
  {Rossat-Mignod}}, \bibinfo {author} {\bibfnamefont {L.~P.}\ \bibnamefont
  {Regnault}}, \bibinfo {author} {\bibfnamefont {C.}~\bibnamefont {Vettier}},
  \bibinfo {author} {\bibfnamefont {P.}~\bibnamefont {Bourges}}, \bibinfo
  {author} {\bibfnamefont {P.}~\bibnamefont {Burlet}}, \bibinfo {author}
  {\bibfnamefont {J.}~\bibnamefont {Bossy}}, \bibinfo {author} {\bibfnamefont
  {J.~Y.}\ \bibnamefont {Henry}}, \ and\ \bibinfo {author} {\bibfnamefont
  {G.}~\bibnamefont {Lapertot}},\ }\bibfield  {title} {\enquote {\bibinfo
  {title} {Neutron scattering study of the {YB}a$_2${C}u$_3${O}$_{6+x}$
  system},}\ }\href
  {http://www.sciencedirect.com/science/article/pii/0921453491919554}
  {\bibfield  {journal} {\bibinfo  {journal} {Physica C}\ }\textbf {\bibinfo
  {volume} {185}},\ \bibinfo {pages} {86} (\bibinfo {year} {1991})}\BibitemShut
  {NoStop}%
\bibitem [{\citenamefont {Mook}\ \emph {et~al.}(1993)\citenamefont {Mook},
  \citenamefont {Yethiraj}, \citenamefont {Aeppli}, \citenamefont {Mason},\
  and\ \citenamefont {Armstrong}}]{Mook_PRL70}%
  \BibitemOpen
  \bibfield  {author} {\bibinfo {author} {\bibfnamefont {H.~A.}\ \bibnamefont
  {Mook}}, \bibinfo {author} {\bibfnamefont {M.}~\bibnamefont {Yethiraj}},
  \bibinfo {author} {\bibfnamefont {G.}~\bibnamefont {Aeppli}}, \bibinfo
  {author} {\bibfnamefont {T.~E.}\ \bibnamefont {Mason}}, \ and\ \bibinfo
  {author} {\bibfnamefont {T.}~\bibnamefont {Armstrong}},\ }\bibfield  {title}
  {\enquote {\bibinfo {title} {Polarized neutron determination of the magnetic
  excitations in {YB}a$_{2}${C}u$_{3}${O}$_{7}$},}\ }\href
  {http://link.aps.org/doi/10.1103/PhysRevLett.70.3490} {\bibfield  {journal}
  {\bibinfo  {journal} {Phys. Rev. Lett.}\ }\textbf {\bibinfo {volume} {70}},\
  \bibinfo {pages} {3490} (\bibinfo {year} {1993})}\BibitemShut {NoStop}%
\bibitem [{\citenamefont {Fong}\ \emph {et~al.}(1999)\citenamefont {Fong},
  \citenamefont {Bourges}, \citenamefont {Sidis}, \citenamefont {Regnault},
  \citenamefont {Ivanov}, \citenamefont {Gu}, \citenamefont {Koshizuka},\ and\
  \citenamefont {Keimer}}]{Fong_Nature398}%
  \BibitemOpen
  \bibfield  {author} {\bibinfo {author} {\bibfnamefont {H.~F.}\ \bibnamefont
  {Fong}}, \bibinfo {author} {\bibfnamefont {P.}~\bibnamefont {Bourges}},
  \bibinfo {author} {\bibfnamefont {Y.}~\bibnamefont {Sidis}}, \bibinfo
  {author} {\bibfnamefont {L.~P.}\ \bibnamefont {Regnault}}, \bibinfo {author}
  {\bibfnamefont {A.}~\bibnamefont {Ivanov}}, \bibinfo {author} {\bibfnamefont
  {G.~D.}\ \bibnamefont {Gu}}, \bibinfo {author} {\bibfnamefont
  {N.}~\bibnamefont {Koshizuka}}, \ and\ \bibinfo {author} {\bibfnamefont
  {B.}~\bibnamefont {Keimer}},\ }\bibfield  {title} {\enquote {\bibinfo {title}
  {Neutron scattering from magnetic excitations in
  {B}i$_2${S}r$_2${C}a{C}u$_2${O}$_{8+\delta}$},}\ }\href
  {http://dx.doi.org/10.1038/19255} {\bibfield  {journal} {\bibinfo  {journal}
  {Nature}\ }\textbf {\bibinfo {volume} {398}},\ \bibinfo {pages} {588}
  (\bibinfo {year} {1999})}\BibitemShut {NoStop}%
\bibitem [{\citenamefont {Dai}\ \emph {et~al.}(2000)\citenamefont {Dai},
  \citenamefont {Mook}, \citenamefont {Aeppli}, \citenamefont {Hayden},\ and\
  \citenamefont {Dogan}}]{Dai_Nature406}%
  \BibitemOpen
  \bibfield  {author} {\bibinfo {author} {\bibfnamefont {P.~C.}\ \bibnamefont
  {Dai}}, \bibinfo {author} {\bibfnamefont {H.~A.}\ \bibnamefont {Mook}},
  \bibinfo {author} {\bibfnamefont {G.}~\bibnamefont {Aeppli}}, \bibinfo
  {author} {\bibfnamefont {S.~M.}\ \bibnamefont {Hayden}}, \ and\ \bibinfo
  {author} {\bibfnamefont {F.}~\bibnamefont {Dogan}},\ }\bibfield  {title}
  {\enquote {\bibinfo {title} {Resonance as a measure of pairing correlations
  in the high-${T}_c$ superconductor {YB}a$_2${C}u$_3${O}$_{6.6}$},}\ }\href
  {\doibase 10.1038/35023094} {\bibfield  {journal} {\bibinfo  {journal}
  {Nature}\ }\textbf {\bibinfo {volume} {406}},\ \bibinfo {pages} {965}
  (\bibinfo {year} {2000})}\BibitemShut {NoStop}%
\bibitem [{\citenamefont {He}\ \emph {et~al.}(2002)\citenamefont {He},
  \citenamefont {Bourges}, \citenamefont {Sidis}, \citenamefont {Ulrich},
  \citenamefont {Regnault}, \citenamefont {Pailh{\`e}s}, \citenamefont
  {Berzigiarova}, \citenamefont {Kolesnikov},\ and\ \citenamefont
  {Keimer}}]{Dai_Science295}%
  \BibitemOpen
  \bibfield  {author} {\bibinfo {author} {\bibfnamefont {H.}~\bibnamefont
  {He}}, \bibinfo {author} {\bibfnamefont {P.}~\bibnamefont {Bourges}},
  \bibinfo {author} {\bibfnamefont {Y.}~\bibnamefont {Sidis}}, \bibinfo
  {author} {\bibfnamefont {C.}~\bibnamefont {Ulrich}}, \bibinfo {author}
  {\bibfnamefont {L.~P.}\ \bibnamefont {Regnault}}, \bibinfo {author}
  {\bibfnamefont {S.}~\bibnamefont {Pailh{\`e}s}}, \bibinfo {author}
  {\bibfnamefont {N.~S.}\ \bibnamefont {Berzigiarova}}, \bibinfo {author}
  {\bibfnamefont {N.~N.}\ \bibnamefont {Kolesnikov}}, \ and\ \bibinfo {author}
  {\bibfnamefont {B.}~\bibnamefont {Keimer}},\ }\bibfield  {title} {\enquote
  {\bibinfo {title} {Magnetic resonant mode in the single-layer
  high-temperature superconductor {T}l$_2${B}a$_2${C}u{O}$_{6+\delta}$},}\
  }\href {\doibase 10.1126/science.1067877} {\bibfield  {journal} {\bibinfo
  {journal} {Science}\ }\textbf {\bibinfo {volume} {295}},\ \bibinfo {pages}
  {1045} (\bibinfo {year} {2002})}\BibitemShut {NoStop}%
\bibitem [{\citenamefont {Eschrig}(2006)}]{Eschrig_AdvPhys55}%
  \BibitemOpen
  \bibfield  {author} {\bibinfo {author} {\bibfnamefont {M.}~\bibnamefont
  {Eschrig}},\ }\bibfield  {title} {\enquote {\bibinfo {title} {The effect of
  collective spin-1 excitations on electronic spectra in high-${T}_c$
  superconductors},}\ }\href {\doibase 10.1080/00018730600645636} {\bibfield
  {journal} {\bibinfo  {journal} {Adv. Phys.}\ }\textbf {\bibinfo {volume}
  {55}},\ \bibinfo {pages} {47--183} (\bibinfo {year} {2006})}\BibitemShut
  {NoStop}%
\bibitem [{\citenamefont {Zhao}\ \emph {et~al.}(2007)\citenamefont {Zhao},
  \citenamefont {Dai}, \citenamefont {Li}, \citenamefont {Freeman},
  \citenamefont {Onose},\ and\ \citenamefont {Tokura}}]{Zhao_PRL99}%
  \BibitemOpen
  \bibfield  {author} {\bibinfo {author} {\bibfnamefont {J.}~\bibnamefont
  {Zhao}}, \bibinfo {author} {\bibfnamefont {P.~C.}\ \bibnamefont {Dai}},
  \bibinfo {author} {\bibfnamefont {S.~L.}\ \bibnamefont {Li}}, \bibinfo
  {author} {\bibfnamefont {P.~G.}\ \bibnamefont {Freeman}}, \bibinfo {author}
  {\bibfnamefont {Y.}~\bibnamefont {Onose}}, \ and\ \bibinfo {author}
  {\bibfnamefont {Y.}~\bibnamefont {Tokura}},\ }\bibfield  {title} {\enquote
  {\bibinfo {title} {Neutron-spin resonance in the optimally electron-doped
  superconductor {N}d$_{1.85}${C}e$_{0.15}${C}u{O}$_{4-\delta}$},}\ }\href
  {\doibase 10.1103/PhysRevLett.99.017001} {\bibfield  {journal} {\bibinfo
  {journal} {Phys. Rev. Lett.}\ }\textbf {\bibinfo {volume} {99}},\ \bibinfo
  {pages} {017001} (\bibinfo {year} {2007})}\BibitemShut {NoStop}%
\bibitem [{\citenamefont {Fong}\ \emph {et~al.}(1995)\citenamefont {Fong},
  \citenamefont {Keimer}, \citenamefont {Anderson}, \citenamefont {Reznik},
  \citenamefont {Do\u{g}an},\ and\ \citenamefont {Aksay}}]{Fong1995PRL}%
  \BibitemOpen
  \bibfield  {author} {\bibinfo {author} {\bibfnamefont {H.~F.}\ \bibnamefont
  {Fong}}, \bibinfo {author} {\bibfnamefont {B.}~\bibnamefont {Keimer}},
  \bibinfo {author} {\bibfnamefont {P.~W.}\ \bibnamefont {Anderson}}, \bibinfo
  {author} {\bibfnamefont {D.}~\bibnamefont {Reznik}}, \bibinfo {author}
  {\bibfnamefont {F.}~\bibnamefont {Do\u{g}an}}, \ and\ \bibinfo {author}
  {\bibfnamefont {I.~A.}\ \bibnamefont {Aksay}},\ }\bibfield  {title} {\enquote
  {\bibinfo {title} {{Phonon and magnetic neutron scattering at 41 meV in
  {Y}{B}${\mathrm{a}}_{2}${C}${\mathrm{u}}_{3}$${\mathrm{{O}}}_{7}$}},}\ }\href
  {\doibase 10.1103/PhysRevLett.75.316} {\bibfield  {journal} {\bibinfo
  {journal} {Phys. Rev. Lett.}\ }\textbf {\bibinfo {volume} {75}},\ \bibinfo
  {pages} {316} (\bibinfo {year} {1995})}\BibitemShut {NoStop}%
\bibitem [{\citenamefont {Blumberg}\ \emph {et~al.}(1995)\citenamefont
  {Blumberg}, \citenamefont {Stojkovi\ifmmode~\acute{c}\else \'{c}\fi{}},\ and\
  \citenamefont {Klein}}]{Blumberg1995PRB}%
  \BibitemOpen
  \bibfield  {author} {\bibinfo {author} {\bibfnamefont {G.}~\bibnamefont
  {Blumberg}}, \bibinfo {author} {\bibfnamefont {Branko~P.}\ \bibnamefont
  {Stojkovi\ifmmode~\acute{c}\else \'{c}\fi{}}}, \ and\ \bibinfo {author}
  {\bibfnamefont {M.~V.}\ \bibnamefont {Klein}},\ }\bibfield  {title} {\enquote
  {\bibinfo {title} {{Antiferromagnetic excitations and van {H}ove
  singularities in
  ${\mathrm{YBa}}_{2}$${\mathrm{Cu}}_{3}$${\mathrm{O}}_{6+\mathit{x}}$}},}\
  }\href {\doibase 10.1103/PhysRevB.52.R15741} {\bibfield  {journal} {\bibinfo
  {journal} {Phys. Rev. B}\ }\textbf {\bibinfo {volume} {52}},\ \bibinfo
  {pages} {R15741} (\bibinfo {year} {1995})}\BibitemShut {NoStop}%
\bibitem [{\citenamefont {Christianson}\ \emph {et~al.}(2008)\citenamefont
  {Christianson}, \citenamefont {Goremychkin}, \citenamefont {Osborn},
  \citenamefont {Rosenkranz}, \citenamefont {Lumsden}, \citenamefont
  {Malliakas}, \citenamefont {Todorov}, \citenamefont {Claus}, \citenamefont
  {Chung}, \citenamefont {Kanatzidis}, \citenamefont {Bewley},\ and\
  \citenamefont {Guidi}}]{christianson2008NatureINS}%
  \BibitemOpen
  \bibfield  {author} {\bibinfo {author} {\bibfnamefont {A.~D.}\ \bibnamefont
  {Christianson}}, \bibinfo {author} {\bibfnamefont {E.~A.}\ \bibnamefont
  {Goremychkin}}, \bibinfo {author} {\bibfnamefont {R.}~\bibnamefont {Osborn}},
  \bibinfo {author} {\bibfnamefont {S.}~\bibnamefont {Rosenkranz}}, \bibinfo
  {author} {\bibfnamefont {M.~D.}\ \bibnamefont {Lumsden}}, \bibinfo {author}
  {\bibfnamefont {C.~D.}\ \bibnamefont {Malliakas}}, \bibinfo {author}
  {\bibfnamefont {I.~S.}\ \bibnamefont {Todorov}}, \bibinfo {author}
  {\bibfnamefont {H.}~\bibnamefont {Claus}}, \bibinfo {author} {\bibfnamefont
  {D.~Y.}\ \bibnamefont {Chung}}, \bibinfo {author} {\bibfnamefont {M.~G.}\
  \bibnamefont {Kanatzidis}}, \bibinfo {author} {\bibfnamefont {R.~I.}\
  \bibnamefont {Bewley}}, \ and\ \bibinfo {author} {\bibfnamefont
  {T.}~\bibnamefont {Guidi}},\ }\bibfield  {title} {\enquote {\bibinfo {title}
  {Unconventional superconductivity in {B}a$_{0.6}${K}$_{0.4}${F}e$_2${A}s$_2$
  from inelastic neutron scattering},}\ }\href {\doibase 10.1038/nature07625}
  {\bibfield  {journal} {\bibinfo  {journal} {Nature}\ }\textbf {\bibinfo
  {volume} {456}},\ \bibinfo {pages} {930} (\bibinfo {year}
  {2008})}\BibitemShut {NoStop}%
\bibitem [{\citenamefont {Zhang}\ \emph {et~al.}(2011)\citenamefont {Zhang},
  \citenamefont {Wang}, \citenamefont {Luo}, \citenamefont {Wang},
  \citenamefont {Liu}, \citenamefont {Zhao}, \citenamefont {Abernathy},
  \citenamefont {Maier}, \citenamefont {Marty}, \citenamefont {Lumsden},
  \citenamefont {Chi}, \citenamefont {Chang}, \citenamefont {Rodriguez-Rivera},
  \citenamefont {Lynn}, \citenamefont {Xiang}, \citenamefont {Hu},\ and\
  \citenamefont {Dai}}]{ZhangCL_SciRep1}%
  \BibitemOpen
  \bibfield  {author} {\bibinfo {author} {\bibfnamefont {C.~L.}\ \bibnamefont
  {Zhang}}, \bibinfo {author} {\bibfnamefont {M.}~\bibnamefont {Wang}},
  \bibinfo {author} {\bibfnamefont {H.~Q.}\ \bibnamefont {Luo}}, \bibinfo
  {author} {\bibfnamefont {M.~Y.}\ \bibnamefont {Wang}}, \bibinfo {author}
  {\bibfnamefont {M.~S.}\ \bibnamefont {Liu}}, \bibinfo {author} {\bibfnamefont
  {J.}~\bibnamefont {Zhao}}, \bibinfo {author} {\bibfnamefont {D.~L.}\
  \bibnamefont {Abernathy}}, \bibinfo {author} {\bibfnamefont {T.~A.}\
  \bibnamefont {Maier}}, \bibinfo {author} {\bibfnamefont {K.}~\bibnamefont
  {Marty}}, \bibinfo {author} {\bibfnamefont {M.~D.}\ \bibnamefont {Lumsden}},
  \bibinfo {author} {\bibfnamefont {S.~X}\ \bibnamefont {Chi}}, \bibinfo
  {author} {\bibfnamefont {S.}~\bibnamefont {Chang}}, \bibinfo {author}
  {\bibfnamefont {J.~A.}\ \bibnamefont {Rodriguez-Rivera}}, \bibinfo {author}
  {\bibfnamefont {J.~W.}\ \bibnamefont {Lynn}}, \bibinfo {author}
  {\bibfnamefont {T.}~\bibnamefont {Xiang}}, \bibinfo {author} {\bibfnamefont
  {J.~P.}\ \bibnamefont {Hu}}, \ and\ \bibinfo {author} {\bibfnamefont {P.~C.}\
  \bibnamefont {Dai}},\ }\bibfield  {title} {\enquote {\bibinfo {title}
  {Neutron scattering studies of spin excitations in hole-doped
  {B}a$_{0.67}${K}$_{0.33}${F}e$_2${A}s$_2$ superconductor},}\ }\href
  {http://dx.doi.org/10.1038/srep00115} {\bibfield  {journal} {\bibinfo
  {journal} {Sci. Rep.}\ }\textbf {\bibinfo {volume} {1}},\ \bibinfo {pages}
  {115} (\bibinfo {year} {2011})}\BibitemShut {NoStop}%
\bibitem [{\citenamefont {Richard}\ \emph {et~al.}(2009)\citenamefont
  {Richard}, \citenamefont {Sato}, \citenamefont {Nakayama}, \citenamefont
  {Souma}, \citenamefont {Takahashi}, \citenamefont {Xu}, \citenamefont {Chen},
  \citenamefont {Luo}, \citenamefont {Wang},\ and\ \citenamefont
  {Ding}}]{Richard2009PRLkink}%
  \BibitemOpen
  \bibfield  {author} {\bibinfo {author} {\bibfnamefont {P.}~\bibnamefont
  {Richard}}, \bibinfo {author} {\bibfnamefont {T.}~\bibnamefont {Sato}},
  \bibinfo {author} {\bibfnamefont {K.}~\bibnamefont {Nakayama}}, \bibinfo
  {author} {\bibfnamefont {S.}~\bibnamefont {Souma}}, \bibinfo {author}
  {\bibfnamefont {T.}~\bibnamefont {Takahashi}}, \bibinfo {author}
  {\bibfnamefont {Y.~M.}\ \bibnamefont {Xu}}, \bibinfo {author} {\bibfnamefont
  {G.~F.}\ \bibnamefont {Chen}}, \bibinfo {author} {\bibfnamefont {J.~L.}\
  \bibnamefont {Luo}}, \bibinfo {author} {\bibfnamefont {N.~L.}\ \bibnamefont
  {Wang}}, \ and\ \bibinfo {author} {\bibfnamefont {H.}~\bibnamefont {Ding}},\
  }\bibfield  {title} {\enquote {\bibinfo {title} {{Angle-resolved
  photoemission spectroscopy of the Fe-Based
  {B}a$_{0.6}${K}$_{0.4}${F}e$_2${A}s$_2$ high temperature superconductor:
  Evidence for an orbital selective electron-mode coupling}},}\ }\href
  {\doibase 10.1103/PhysRevLett.102.047003} {\bibfield  {journal} {\bibinfo
  {journal} {Phys. Rev. Lett.}\ }\textbf {\bibinfo {volume} {102}},\ \bibinfo
  {pages} {047003} (\bibinfo {year} {2009})}\BibitemShut {NoStop}%
\bibitem [{\citenamefont {Shan}\ \emph {et~al.}(2012)\citenamefont {Shan},
  \citenamefont {Gong}, \citenamefont {Wang}, \citenamefont {Shen},
  \citenamefont {Hou}, \citenamefont {Ren}, \citenamefont {Li}, \citenamefont
  {Yang}, \citenamefont {Wen}, \citenamefont {Li},\ and\ \citenamefont
  {Dai}}]{shan2012PRLSTM}%
  \BibitemOpen
  \bibfield  {author} {\bibinfo {author} {\bibfnamefont {L.}~\bibnamefont
  {Shan}}, \bibinfo {author} {\bibfnamefont {J.}~\bibnamefont {Gong}}, \bibinfo
  {author} {\bibfnamefont {Y.~L.}\ \bibnamefont {Wang}}, \bibinfo {author}
  {\bibfnamefont {B.}~\bibnamefont {Shen}}, \bibinfo {author} {\bibfnamefont
  {X.~Y.}\ \bibnamefont {Hou}}, \bibinfo {author} {\bibfnamefont
  {C.}~\bibnamefont {Ren}}, \bibinfo {author} {\bibfnamefont {C.H}\
  \bibnamefont {Li}}, \bibinfo {author} {\bibfnamefont {H.}~\bibnamefont
  {Yang}}, \bibinfo {author} {\bibfnamefont {H.~H.}\ \bibnamefont {Wen}},
  \bibinfo {author} {\bibfnamefont {S.~L.}\ \bibnamefont {Li}}, \ and\ \bibinfo
  {author} {\bibfnamefont {P.~C}\ \bibnamefont {Dai}},\ }\bibfield  {title}
  {\enquote {\bibinfo {title} {Evidence of a spin resonance mode in the
  iron-based superconductor {B}a$_{0.6}${K}$_{0.4}${F}e$_2${A}s$_2$ from
  scanning tunneling spectroscopy},}\ }\href
  {http://link.aps.org/doi/10.1103/PhysRevLett.108.227002} {\bibfield
  {journal} {\bibinfo  {journal} {Phys. Rev. Lett.}\ }\textbf {\bibinfo
  {volume} {108}},\ \bibinfo {pages} {227002} (\bibinfo {year}
  {2012})}\BibitemShut {NoStop}%
\bibitem [{\citenamefont {Wu}\ \emph {et~al.}(2010)\citenamefont {Wu},
  \citenamefont {Bari\ifmmode \check{s}\else \v{s}\fi{}i\ifmmode~\acute{c}\else
  \'{c}\fi{}}, \citenamefont {Dressel}, \citenamefont {Cao}, \citenamefont
  {Xu}, \citenamefont {Schachinger},\ and\ \citenamefont
  {Carbotte}}]{Wu_PRB82}%
  \BibitemOpen
  \bibfield  {author} {\bibinfo {author} {\bibfnamefont {D.}~\bibnamefont
  {Wu}}, \bibinfo {author} {\bibfnamefont {N.}~\bibnamefont {Bari\ifmmode
  \check{s}\else \v{s}\fi{}i\ifmmode~\acute{c}\else \'{c}\fi{}}}, \bibinfo
  {author} {\bibfnamefont {M.}~\bibnamefont {Dressel}}, \bibinfo {author}
  {\bibfnamefont {G.~H.}\ \bibnamefont {Cao}}, \bibinfo {author} {\bibfnamefont
  {Z-A.}\ \bibnamefont {Xu}}, \bibinfo {author} {\bibfnamefont
  {E.}~\bibnamefont {Schachinger}}, \ and\ \bibinfo {author} {\bibfnamefont
  {J.~P.}\ \bibnamefont {Carbotte}},\ }\bibfield  {title} {\enquote {\bibinfo
  {title} {Eliashberg analysis of optical spectra reveals a strong coupling of
  charge carriers to spin fluctuations in doped iron-pnictide
  {B}a{F}e$_{2}${A}s$_2$ superconductors},}\ }\href {\doibase
  10.1103/PhysRevB.82.144519} {\bibfield  {journal} {\bibinfo  {journal} {Phys.
  Rev. B}\ }\textbf {\bibinfo {volume} {82}},\ \bibinfo {pages} {144519}
  (\bibinfo {year} {2010})}\BibitemShut {NoStop}%
\bibitem [{\citenamefont {Fernandes}\ \emph {et~al.}(2014)\citenamefont
  {Fernandes}, \citenamefont {Chubukov},\ and\ \citenamefont
  {Schmalian}}]{Fernandes2014NatPhy}%
  \BibitemOpen
  \bibfield  {author} {\bibinfo {author} {\bibfnamefont {R.~M.}\ \bibnamefont
  {Fernandes}}, \bibinfo {author} {\bibfnamefont {A.~V.}\ \bibnamefont
  {Chubukov}}, \ and\ \bibinfo {author} {\bibfnamefont {J.}~\bibnamefont
  {Schmalian}},\ }\bibfield  {title} {\enquote {\bibinfo {title} {What drives
  nematic order in iron-based superconductors?}}\ }\href
  {http://dx.doi.org/10.1038/nphys2877} {\bibfield  {journal} {\bibinfo
  {journal} {Nature Phys.}\ }\textbf {\bibinfo {volume} {10}},\ \bibinfo
  {pages} {97} (\bibinfo {year} {2014})}\BibitemShut {NoStop}%
\bibitem [{\citenamefont {Thorsm\o{}lle}\ \emph {et~al.}(2016)\citenamefont
  {Thorsm\o{}lle}, \citenamefont {Khodas}, \citenamefont {Yin}, \citenamefont
  {Zhang}, \citenamefont {Carr}, \citenamefont {Dai},\ and\ \citenamefont
  {Blumberg}}]{Thorsmolle2016PRB}%
  \BibitemOpen
  \bibfield  {author} {\bibinfo {author} {\bibfnamefont {V.~K.}\ \bibnamefont
  {Thorsm\o{}lle}}, \bibinfo {author} {\bibfnamefont {M.}~\bibnamefont
  {Khodas}}, \bibinfo {author} {\bibfnamefont {Z.~P.}\ \bibnamefont {Yin}},
  \bibinfo {author} {\bibfnamefont {C.~L.}\ \bibnamefont {Zhang}}, \bibinfo
  {author} {\bibfnamefont {S.~V.}\ \bibnamefont {Carr}}, \bibinfo {author}
  {\bibfnamefont {P.~C}\ \bibnamefont {Dai}}, \ and\ \bibinfo {author}
  {\bibfnamefont {G.}~\bibnamefont {Blumberg}},\ }\bibfield  {title} {\enquote
  {\bibinfo {title} {Critical quadrupole fluctuations and collective modes in
  iron pnictide superconductors},}\ }\href
  {http://link.aps.org/doi/10.1103/PhysRevB.93.054515} {\bibfield  {journal}
  {\bibinfo  {journal} {Phys. Rev. B}\ }\textbf {\bibinfo {volume} {93}},\
  \bibinfo {pages} {054515} (\bibinfo {year} {2016})}\BibitemShut {NoStop}%
\bibitem [{\citenamefont {Gallais}\ \emph {et~al.}(2016)\citenamefont
  {Gallais}, \citenamefont {Paul}, \citenamefont {Chauviere},\ and\
  \citenamefont {Schmalian}}]{Gallais2016PRL}%
  \BibitemOpen
  \bibfield  {author} {\bibinfo {author} {\bibfnamefont {Y.}~\bibnamefont
  {Gallais}}, \bibinfo {author} {\bibfnamefont {I.}~\bibnamefont {Paul}},
  \bibinfo {author} {\bibfnamefont {L.}~\bibnamefont {Chauviere}}, \ and\
  \bibinfo {author} {\bibfnamefont {J.}~\bibnamefont {Schmalian}},\ }\bibfield
  {title} {\enquote {\bibinfo {title} {Nematic resonance in the {R}aman
  response of iron-based superconductors},}\ }\href
  {http://link.aps.org/doi/10.1103/PhysRevLett.116.017001} {\bibfield
  {journal} {\bibinfo  {journal} {Phys. Rev. Lett.}\ }\textbf {\bibinfo
  {volume} {116}},\ \bibinfo {pages} {017001} (\bibinfo {year}
  {2016})}\BibitemShut {NoStop}%
\bibitem [{\citenamefont {Kretzschmar}\ \emph {et~al.}(2013)\citenamefont
  {Kretzschmar}, \citenamefont {Muschler}, \citenamefont {B\"ohm},
  \citenamefont {Baum}, \citenamefont {Hackl}, \citenamefont {Wen},
  \citenamefont {Tsurkan}, \citenamefont {Deisenhofer},\ and\ \citenamefont
  {Loidl}}]{Rudi_2013PRL}%
  \BibitemOpen
  \bibfield  {author} {\bibinfo {author} {\bibfnamefont {F.}~\bibnamefont
  {Kretzschmar}}, \bibinfo {author} {\bibfnamefont {B.}~\bibnamefont
  {Muschler}}, \bibinfo {author} {\bibfnamefont {T.}~\bibnamefont {B\"ohm}},
  \bibinfo {author} {\bibfnamefont {A.}~\bibnamefont {Baum}}, \bibinfo {author}
  {\bibfnamefont {R.}~\bibnamefont {Hackl}}, \bibinfo {author} {\bibfnamefont
  {H.~H}\ \bibnamefont {Wen}}, \bibinfo {author} {\bibfnamefont
  {V.}~\bibnamefont {Tsurkan}}, \bibinfo {author} {\bibfnamefont
  {J.}~\bibnamefont {Deisenhofer}}, \ and\ \bibinfo {author} {\bibfnamefont
  {A.}~\bibnamefont {Loidl}},\ }\bibfield  {title} {\enquote {\bibinfo {title}
  {{R}aman-scattering detection of nearly degenerate $s$-wave and $d$-wave
  pairing channels in iron-based {B}a$_{0.6}${K}$_{0.4}${F}e$_2${A}s$_2$ and
  {R}b$_{0.8}${F}e$_{1.6}${S}e$_2$ superconductors},}\ }\href
  {http://link.aps.org/doi/10.1103/PhysRevLett.110.187002} {\bibfield
  {journal} {\bibinfo  {journal} {Phys. Rev. Lett.}\ }\textbf {\bibinfo
  {volume} {110}},\ \bibinfo {pages} {187002} (\bibinfo {year}
  {2013})}\BibitemShut {NoStop}%
\bibitem [{\citenamefont {B\"ohm}\ \emph {et~al.}(2014)\citenamefont {B\"ohm},
  \citenamefont {Kemper}, \citenamefont {Moritz}, \citenamefont {Kretzschmar},
  \citenamefont {Muschler}, \citenamefont {Eiter}, \citenamefont {Hackl},
  \citenamefont {Devereaux}, \citenamefont {Scalapino},\ and\ \citenamefont
  {Wen}}]{Rudi_2014PRX}%
  \BibitemOpen
  \bibfield  {author} {\bibinfo {author} {\bibfnamefont {T.}~\bibnamefont
  {B\"ohm}}, \bibinfo {author} {\bibfnamefont {A.~F.}\ \bibnamefont {Kemper}},
  \bibinfo {author} {\bibfnamefont {B.}~\bibnamefont {Moritz}}, \bibinfo
  {author} {\bibfnamefont {F.}~\bibnamefont {Kretzschmar}}, \bibinfo {author}
  {\bibfnamefont {B.}~\bibnamefont {Muschler}}, \bibinfo {author}
  {\bibfnamefont {H.-M.}\ \bibnamefont {Eiter}}, \bibinfo {author}
  {\bibfnamefont {R.}~\bibnamefont {Hackl}}, \bibinfo {author} {\bibfnamefont
  {T.~P.}\ \bibnamefont {Devereaux}}, \bibinfo {author} {\bibfnamefont {D.~J.}\
  \bibnamefont {Scalapino}}, \ and\ \bibinfo {author} {\bibfnamefont {H.~H}\
  \bibnamefont {Wen}},\ }\bibfield  {title} {\enquote {\bibinfo {title}
  {Balancing act: Evidence for a strong subdominant $d$-wave pairing channel in
  {B}a$_{0.6}${K}$_{0.4}${F}e$_2${A}s$_2$},}\ }\href
  {http://link.aps.org/doi/10.1103/PhysRevX.4.041046} {\bibfield  {journal}
  {\bibinfo  {journal} {Phys. Rev. X}\ }\textbf {\bibinfo {volume} {4}},\
  \bibinfo {pages} {041046} (\bibinfo {year} {2014})}\BibitemShut {NoStop}%
\bibitem [{\citenamefont {B{\"o}hm}\ \emph {et~al.}(2016)\citenamefont
  {B{\"o}hm}, \citenamefont {Hosseinian~Ahangharnejhad}, \citenamefont {Jost},
  \citenamefont {Baum}, \citenamefont {Muschler}, \citenamefont {Kretzschmar},
  \citenamefont {Adelmann}, \citenamefont {Wolf}, \citenamefont {Wen},
  \citenamefont {Chu}, \citenamefont {Fisher},\ and\ \citenamefont
  {Hackl}}]{bohm2016PSSB}%
  \BibitemOpen
  \bibfield  {author} {\bibinfo {author} {\bibfnamefont {T.}~\bibnamefont
  {B{\"o}hm}}, \bibinfo {author} {\bibfnamefont {R.}~\bibnamefont
  {Hosseinian~Ahangharnejhad}}, \bibinfo {author} {\bibfnamefont
  {D.}~\bibnamefont {Jost}}, \bibinfo {author} {\bibfnamefont {A.}~\bibnamefont
  {Baum}}, \bibinfo {author} {\bibfnamefont {B.}~\bibnamefont {Muschler}},
  \bibinfo {author} {\bibfnamefont {F.}~\bibnamefont {Kretzschmar}}, \bibinfo
  {author} {\bibfnamefont {P.}~\bibnamefont {Adelmann}}, \bibinfo {author}
  {\bibfnamefont {T.}~\bibnamefont {Wolf}}, \bibinfo {author} {\bibfnamefont
  {H.~H.}\ \bibnamefont {Wen}}, \bibinfo {author} {\bibfnamefont {J.~H.}\
  \bibnamefont {Chu}}, \bibinfo {author} {\bibfnamefont {I.~R.}\ \bibnamefont
  {Fisher}}, \ and\ \bibinfo {author} {\bibfnamefont {R.}~\bibnamefont
  {Hackl}},\ }\bibfield  {title} {\enquote {\bibinfo {title} {Superconductivity
  and fluctuations in {B}a$_{1-p}${K}$_p${F}e$_2${A}s$_2$ and
  {B}a({F}e$_{1-n}${C}o$_n$)$_2${A}s$_2$},}\ }\href
  {http://dx.doi.org/10.1002/pssb.201600308} {\bibfield  {journal} {\bibinfo
  {journal} {Phys. Status Solidi B}\ ,\ \bibinfo {pages} {DOI
  10.1002/pssb.201600308}} (\bibinfo {year} {2016})}\BibitemShut {NoStop}%
\bibitem [{\citenamefont {Bardasis}\ and\ \citenamefont
  {Schrieffer}(1961)}]{BS1961PR}%
  \BibitemOpen
  \bibfield  {author} {\bibinfo {author} {\bibfnamefont {A.}~\bibnamefont
  {Bardasis}}\ and\ \bibinfo {author} {\bibfnamefont {J.~R.}\ \bibnamefont
  {Schrieffer}},\ }\bibfield  {title} {\enquote {\bibinfo {title} {Excitons and
  plasmons in superconductors},}\ }\href
  {http://link.aps.org/doi/10.1103/PhysRev.121.1050} {\bibfield  {journal}
  {\bibinfo  {journal} {Phys. Rev.}\ }\textbf {\bibinfo {volume} {121}},\
  \bibinfo {pages} {1050} (\bibinfo {year} {1961})}\BibitemShut {NoStop}%
\bibitem [{\citenamefont {Tsuneto}(1960)}]{Tsuneto1960PR}%
  \BibitemOpen
  \bibfield  {author} {\bibinfo {author} {\bibfnamefont {T.}~\bibnamefont
  {Tsuneto}},\ }\bibfield  {title} {\enquote {\bibinfo {title} {Transverse
  collective excitations in superconductors and electromagnetic absorption},}\
  }\href {\doibase 10.1103/PhysRev.118.1029} {\bibfield  {journal} {\bibinfo
  {journal} {Phys. Rev.}\ }\textbf {\bibinfo {volume} {118}},\ \bibinfo {pages}
  {1029} (\bibinfo {year} {1960})}\BibitemShut {NoStop}%
\bibitem [{\citenamefont {Klein}\ and\ \citenamefont
  {Dierker}(1984)}]{Klein1984}%
  \BibitemOpen
  \bibfield  {author} {\bibinfo {author} {\bibfnamefont {M.~V.}\ \bibnamefont
  {Klein}}\ and\ \bibinfo {author} {\bibfnamefont {S.~B.}\ \bibnamefont
  {Dierker}},\ }\bibfield  {title} {\enquote {\bibinfo {title} {Theory of
  {R}aman scattering in superconductors},}\ }\href
  {http://link.aps.org/doi/10.1103/PhysRevB.29.4976} {\bibfield  {journal}
  {\bibinfo  {journal} {Phys. Rev. B}\ }\textbf {\bibinfo {volume} {29}},\
  \bibinfo {pages} {4976} (\bibinfo {year} {1984})}\BibitemShut {NoStop}%
\bibitem [{\citenamefont {Klein}(2010)}]{Klein_PRB82}%
  \BibitemOpen
  \bibfield  {author} {\bibinfo {author} {\bibfnamefont {M.~V.}\ \bibnamefont
  {Klein}},\ }\bibfield  {title} {\enquote {\bibinfo {title} {Theory of {R}aman
  scattering from {L}eggett's collective mode in a multiband superconductor:
  Application to {M}g{B}$_{2}$},}\ }\href {\doibase 10.1103/PhysRevB.82.014507}
  {\bibfield  {journal} {\bibinfo  {journal} {Phys. Rev. B}\ }\textbf {\bibinfo
  {volume} {82}},\ \bibinfo {pages} {014507} (\bibinfo {year}
  {2010})}\BibitemShut {NoStop}%
\bibitem [{\citenamefont {Lee}\ \emph {et~al.}(2009)\citenamefont {Lee},
  \citenamefont {Zhang},\ and\ \citenamefont {Wu}}]{LeeWC2009PRL}%
  \BibitemOpen
  \bibfield  {author} {\bibinfo {author} {\bibfnamefont {W.~C.}\ \bibnamefont
  {Lee}}, \bibinfo {author} {\bibfnamefont {S.~C.}\ \bibnamefont {Zhang}}, \
  and\ \bibinfo {author} {\bibfnamefont {C.~J.}\ \bibnamefont {Wu}},\
  }\bibfield  {title} {\enquote {\bibinfo {title} {Pairing state with a
  time-reversal symmetry breaking in {F}e{A}s-based superconductors},}\ }\href
  {\doibase 10.1103/PhysRevLett.102.217002} {\bibfield  {journal} {\bibinfo
  {journal} {Phys. Rev. Lett.}\ }\textbf {\bibinfo {volume} {102}},\ \bibinfo
  {pages} {217002} (\bibinfo {year} {2009})}\BibitemShut {NoStop}%
\bibitem [{\citenamefont {Maiti}\ and\ \citenamefont
  {Hirschfeld}(2015)}]{Maiti2015PRB}%
  \BibitemOpen
  \bibfield  {author} {\bibinfo {author} {\bibfnamefont {S.}~\bibnamefont
  {Maiti}}\ and\ \bibinfo {author} {\bibfnamefont {P.~J.}\ \bibnamefont
  {Hirschfeld}},\ }\bibfield  {title} {\enquote {\bibinfo {title} {Collective
  modes in superconductors with competing $s$- and $d$-wave interactions},}\
  }\href {\doibase 10.1103/PhysRevB.92.094506} {\bibfield  {journal} {\bibinfo
  {journal} {Phys. Rev. B}\ }\textbf {\bibinfo {volume} {92}},\ \bibinfo
  {pages} {094506} (\bibinfo {year} {2015})}\BibitemShut {NoStop}%
\bibitem [{\citenamefont {Scalapino}\ and\ \citenamefont
  {Devereaux}(2009)}]{scalapino2009PRB}%
  \BibitemOpen
  \bibfield  {author} {\bibinfo {author} {\bibfnamefont {D.~J.}\ \bibnamefont
  {Scalapino}}\ and\ \bibinfo {author} {\bibfnamefont {T.~P.}\ \bibnamefont
  {Devereaux}},\ }\bibfield  {title} {\enquote {\bibinfo {title} {Collective
  d-wave exciton modes in the calculated raman spectrum of fe-based
  superconductors},}\ }\href {\doibase 10.1103/PhysRevB.80.140512} {\bibfield
  {journal} {\bibinfo  {journal} {Phys. Rev. B}\ }\textbf {\bibinfo {volume}
  {80}},\ \bibinfo {pages} {140512} (\bibinfo {year} {2009})}\BibitemShut
  {NoStop}%
\bibitem [{\citenamefont {Maiti}\ \emph {et~al.}(2016)\citenamefont {Maiti},
  \citenamefont {Maier}, \citenamefont {Boehm}, \citenamefont {Hackl},\ and\
  \citenamefont {Hirschfeld}}]{Maiti2016arxiv}%
  \BibitemOpen
  \bibfield  {author} {\bibinfo {author} {\bibfnamefont {S.}~\bibnamefont
  {Maiti}}, \bibinfo {author} {\bibfnamefont {T.A.}\ \bibnamefont {Maier}},
  \bibinfo {author} {\bibfnamefont {T.}~\bibnamefont {Boehm}}, \bibinfo
  {author} {\bibfnamefont {R.}~\bibnamefont {Hackl}}, \ and\ \bibinfo {author}
  {\bibfnamefont {P.~J.}\ \bibnamefont {Hirschfeld}},\ }\bibfield  {title}
  {\enquote {\bibinfo {title} {Probing the pairing interaction and multiple
  {B}ardasis-{S}chrieffer modes using {R}aman spectroscopy},}\ }\href
  {https://arxiv.org/abs/1611.04541} {\bibfield  {journal} {\bibinfo  {journal}
  {arXiv:1611.04541}\ } (\bibinfo {year} {2016})}\BibitemShut {NoStop}%
\bibitem [{\citenamefont {Chubukov}\ \emph {et~al.}(2009)\citenamefont
  {Chubukov}, \citenamefont {Eremin},\ and\ \citenamefont
  {Korshunov}}]{Chubukov2009PRB}%
  \BibitemOpen
  \bibfield  {author} {\bibinfo {author} {\bibfnamefont {A.~V.}\ \bibnamefont
  {Chubukov}}, \bibinfo {author} {\bibfnamefont {I.}~\bibnamefont {Eremin}}, \
  and\ \bibinfo {author} {\bibfnamefont {M.~M.}\ \bibnamefont {Korshunov}},\
  }\bibfield  {title} {\enquote {\bibinfo {title} {Theory of {R}aman response
  of a superconductor with extended $s$-wave symmetry: Application to the iron
  pnictides},}\ }\href {\doibase 10.1103/PhysRevB.79.220501} {\bibfield
  {journal} {\bibinfo  {journal} {Phys. Rev. B}\ }\textbf {\bibinfo {volume}
  {79}},\ \bibinfo {pages} {220501} (\bibinfo {year} {2009})}\BibitemShut
  {NoStop}%
\bibitem [{\citenamefont {Khodas}\ \emph {et~al.}(2014)\citenamefont {Khodas},
  \citenamefont {Chubukov},\ and\ \citenamefont {Blumberg}}]{Khodas2014PRB}%
  \BibitemOpen
  \bibfield  {author} {\bibinfo {author} {\bibfnamefont {M.}~\bibnamefont
  {Khodas}}, \bibinfo {author} {\bibfnamefont {A.~V.}\ \bibnamefont
  {Chubukov}}, \ and\ \bibinfo {author} {\bibfnamefont {G.}~\bibnamefont
  {Blumberg}},\ }\bibfield  {title} {\enquote {\bibinfo {title} {Collective
  modes in multiband superconductors: {R}aman scattering in iron selenides},}\
  }\href {\doibase 10.1103/PhysRevB.89.245134} {\bibfield  {journal} {\bibinfo
  {journal} {Phys. Rev. B}\ }\textbf {\bibinfo {volume} {89}},\ \bibinfo
  {pages} {245134} (\bibinfo {year} {2014})}\BibitemShut {NoStop}%
\bibitem [{\citenamefont {Hinojosa}\ \emph {et~al.}(2016)\citenamefont
  {Hinojosa}, \citenamefont {Cai},\ and\ \citenamefont
  {Chubukov}}]{Chubukov2016PRB}%
  \BibitemOpen
  \bibfield  {author} {\bibinfo {author} {\bibfnamefont {A}~\bibnamefont
  {Hinojosa}}, \bibinfo {author} {\bibfnamefont {J.~S}\ \bibnamefont {Cai}}, \
  and\ \bibinfo {author} {\bibfnamefont {A.~V.}\ \bibnamefont {Chubukov}},\
  }\bibfield  {title} {\enquote {\bibinfo {title} {{{R}aman} resonance in
  iron-based superconductors: The magnetic scenario},}\ }\href {\doibase
  10.1103/PhysRevB.93.075106} {\bibfield  {journal} {\bibinfo  {journal} {Phys.
  Rev. B}\ }\textbf {\bibinfo {volume} {93}},\ \bibinfo {pages} {075106}
  (\bibinfo {year} {2016})}\BibitemShut {NoStop}%
\bibitem [{\citenamefont {Shen}\ \emph {et~al.}(2011)\citenamefont {Shen},
  \citenamefont {Yang}, \citenamefont {Wang}, \citenamefont {Han},
  \citenamefont {Zeng}, \citenamefont {Shan}, \citenamefont {Ren},\ and\
  \citenamefont {Wen}}]{2011HHWen_PRB84}%
  \BibitemOpen
  \bibfield  {author} {\bibinfo {author} {\bibfnamefont {B.}~\bibnamefont
  {Shen}}, \bibinfo {author} {\bibfnamefont {H.}~\bibnamefont {Yang}}, \bibinfo
  {author} {\bibfnamefont {Z.~S.}\ \bibnamefont {Wang}}, \bibinfo {author}
  {\bibfnamefont {F.}~\bibnamefont {Han}}, \bibinfo {author} {\bibfnamefont
  {B.}~\bibnamefont {Zeng}}, \bibinfo {author} {\bibfnamefont {L.}~\bibnamefont
  {Shan}}, \bibinfo {author} {\bibfnamefont {C.}~\bibnamefont {Ren}}, \ and\
  \bibinfo {author} {\bibfnamefont {H.~H.}\ \bibnamefont {Wen}},\ }\bibfield
  {title} {\enquote {\bibinfo {title} {Transport properties and asymmetric
  scattering in {B}a${}_{1\ensuremath{-}x}${K}${}_{x}${F}e${}_{2}${A}s${}_{2}$
  single crystals},}\ }\href {\doibase 10.1103/PhysRevB.84.184512} {\bibfield
  {journal} {\bibinfo  {journal} {Phys. Rev. B}\ }\textbf {\bibinfo {volume}
  {84}},\ \bibinfo {pages} {184512} (\bibinfo {year} {2011})}\BibitemShut
  {NoStop}%
\bibitem [{\citenamefont {Devereaux}\ and\ \citenamefont
  {Hackl}(2007)}]{Devereaux2007RMP}%
  \BibitemOpen
  \bibfield  {author} {\bibinfo {author} {\bibfnamefont {T.~P.}\ \bibnamefont
  {Devereaux}}\ and\ \bibinfo {author} {\bibfnamefont {R.}~\bibnamefont
  {Hackl}},\ }\bibfield  {title} {\enquote {\bibinfo {title} {Inelastic light
  scattering from correlated electrons},}\ }\href
  {http://link.aps.org/doi/10.1103/RevModPhys.79.175} {\bibfield  {journal}
  {\bibinfo  {journal} {Rev. Mod. Phys.}\ }\textbf {\bibinfo {volume} {79}},\
  \bibinfo {pages} {175} (\bibinfo {year} {2007})}\BibitemShut {NoStop}%
\bibitem [{\citenamefont {B\"ohmer}\ \emph {et~al.}(2014)\citenamefont
  {B\"ohmer}, \citenamefont {Burger}, \citenamefont {Hardy}, \citenamefont
  {Wolf}, \citenamefont {Schweiss}, \citenamefont {Fromknecht}, \citenamefont
  {Reinecker}, \citenamefont {Schranz},\ and\ \citenamefont
  {Meingast}}]{Bohmer2014PRL}%
  \BibitemOpen
  \bibfield  {author} {\bibinfo {author} {\bibfnamefont {A.~E.}\ \bibnamefont
  {B\"ohmer}}, \bibinfo {author} {\bibfnamefont {P.}~\bibnamefont {Burger}},
  \bibinfo {author} {\bibfnamefont {F.}~\bibnamefont {Hardy}}, \bibinfo
  {author} {\bibfnamefont {T.}~\bibnamefont {Wolf}}, \bibinfo {author}
  {\bibfnamefont {P.}~\bibnamefont {Schweiss}}, \bibinfo {author}
  {\bibfnamefont {R.}~\bibnamefont {Fromknecht}}, \bibinfo {author}
  {\bibfnamefont {M.}~\bibnamefont {Reinecker}}, \bibinfo {author}
  {\bibfnamefont {W.}~\bibnamefont {Schranz}}, \ and\ \bibinfo {author}
  {\bibfnamefont {C.}~\bibnamefont {Meingast}},\ }\bibfield  {title} {\enquote
  {\bibinfo {title} {Nematic susceptibility of hole-doped and electron-doped
  {B}a{F}e$_2${A}s$_2$ iron-based superconductors from shear modulus
  measurements},}\ }\href {\doibase 10.1103/PhysRevLett.112.047001} {\bibfield
  {journal} {\bibinfo  {journal} {Phys. Rev. Lett.}\ }\textbf {\bibinfo
  {volume} {112}},\ \bibinfo {pages} {047001} (\bibinfo {year}
  {2014})}\BibitemShut {NoStop}%
\bibitem [{\citenamefont {Rahlenbeck}\ \emph {et~al.}(2009)\citenamefont
  {Rahlenbeck}, \citenamefont {Sun}, \citenamefont {Sun}, \citenamefont {Lin},
  \citenamefont {Keimer},\ and\ \citenamefont {Ulrich}}]{RahlenbeckPRB80}%
  \BibitemOpen
  \bibfield  {author} {\bibinfo {author} {\bibfnamefont {M.}~\bibnamefont
  {Rahlenbeck}}, \bibinfo {author} {\bibfnamefont {G.~L.}\ \bibnamefont {Sun}},
  \bibinfo {author} {\bibfnamefont {D.~L.}\ \bibnamefont {Sun}}, \bibinfo
  {author} {\bibfnamefont {C.~T.}\ \bibnamefont {Lin}}, \bibinfo {author}
  {\bibfnamefont {B.}~\bibnamefont {Keimer}}, \ and\ \bibinfo {author}
  {\bibfnamefont {C.}~\bibnamefont {Ulrich}},\ }\bibfield  {title} {\enquote
  {\bibinfo {title} {Phonon anomalies in pure and underdoped
  {R}$_{1-x}${K}$_{x}${F}e$_{2}${A}s$_{2}$ ({R}={B}a, {S}r) investigated by
  {R}aman light scattering},}\ }\href {\doibase 10.1103/PhysRevB.80.064509}
  {\bibfield  {journal} {\bibinfo  {journal} {Phys. Rev. B}\ }\textbf {\bibinfo
  {volume} {80}},\ \bibinfo {pages} {064509} (\bibinfo {year}
  {2009})}\BibitemShut {NoStop}%
\bibitem [{\citenamefont {Zhang}\ \emph
  {et~al.}(2014{\natexlab{a}})\citenamefont {Zhang}, \citenamefont {Richard},
  \citenamefont {Ding}, \citenamefont {Sefat}, \citenamefont {Gillett},
  \citenamefont {Sebastian}, \citenamefont {Khodas},\ and\ \citenamefont
  {Blumberg}}]{ZhangWL2014arxiv}%
  \BibitemOpen
  \bibfield  {author} {\bibinfo {author} {\bibfnamefont {W.~L.}\ \bibnamefont
  {Zhang}}, \bibinfo {author} {\bibfnamefont {P.}~\bibnamefont {Richard}},
  \bibinfo {author} {\bibfnamefont {H.}~\bibnamefont {Ding}}, \bibinfo {author}
  {\bibfnamefont {A.~S.}\ \bibnamefont {Sefat}}, \bibinfo {author}
  {\bibfnamefont {J.}~\bibnamefont {Gillett}}, \bibinfo {author} {\bibfnamefont
  {S.~E.}\ \bibnamefont {Sebastian}}, \bibinfo {author} {\bibfnamefont
  {M.}~\bibnamefont {Khodas}}, \ and\ \bibinfo {author} {\bibfnamefont
  {G.}~\bibnamefont {Blumberg}},\ }\bibfield  {title} {\enquote {\bibinfo
  {title} {On the origin of the electronic anisotropy in iron pnicitde
  superconductors},}\ }\href {https://arxiv.org/abs/1410.6452} {\bibfield
  {journal} {\bibinfo  {journal} {arXiv:1410.6452}\ } (\bibinfo {year}
  {2014}{\natexlab{a}})}\BibitemShut {NoStop}%
\bibitem [{\citenamefont {Gallais}\ \emph {et~al.}(2013)\citenamefont
  {Gallais}, \citenamefont {Fernandes}, \citenamefont {Paul}, \citenamefont
  {Chauvi\`ere}, \citenamefont {Yang}, \citenamefont {M\'easson}, \citenamefont
  {Cazayous}, \citenamefont {Sacuto}, \citenamefont {Colson},\ and\
  \citenamefont {Forget}}]{Gallais2013PRL}%
  \BibitemOpen
  \bibfield  {author} {\bibinfo {author} {\bibfnamefont {Y.}~\bibnamefont
  {Gallais}}, \bibinfo {author} {\bibfnamefont {R.~M.}\ \bibnamefont
  {Fernandes}}, \bibinfo {author} {\bibfnamefont {I.}~\bibnamefont {Paul}},
  \bibinfo {author} {\bibfnamefont {L.}~\bibnamefont {Chauvi\`ere}}, \bibinfo
  {author} {\bibfnamefont {Y.~X.}\ \bibnamefont {Yang}}, \bibinfo {author}
  {\bibfnamefont {M.~A.}\ \bibnamefont {M\'easson}}, \bibinfo {author}
  {\bibfnamefont {M.}~\bibnamefont {Cazayous}}, \bibinfo {author}
  {\bibfnamefont {A.}~\bibnamefont {Sacuto}}, \bibinfo {author} {\bibfnamefont
  {D.}~\bibnamefont {Colson}}, \ and\ \bibinfo {author} {\bibfnamefont
  {A.}~\bibnamefont {Forget}},\ }\bibfield  {title} {\enquote {\bibinfo {title}
  {Observation of incipient charge nematicity in
  {B}a({F}e$_{1-x}${C}o$_{x}$)$_2${A}s$_2$},}\ }\href
  {http://link.aps.org/doi/10.1103/PhysRevLett.111.267001} {\bibfield
  {journal} {\bibinfo  {journal} {Phys. Rev. Lett.}\ }\textbf {\bibinfo
  {volume} {111}},\ \bibinfo {pages} {267001} (\bibinfo {year}
  {2013})}\BibitemShut {NoStop}%
\bibitem [{\citenamefont {Kretzschmar}\ \emph {et~al.}(2016)\citenamefont
  {Kretzschmar}, \citenamefont {Bohm}, \citenamefont {Karahasanovic},
  \citenamefont {Muschler}, \citenamefont {Baum}, \citenamefont {Jost},
  \citenamefont {Schmalian}, \citenamefont {Caprara}, \citenamefont {Grilli},
  \citenamefont {Di~Castro}, \citenamefont {Analytis}, \citenamefont {Chu},
  \citenamefont {Fisher},\ and\ \citenamefont {Hackl}}]{Kretzschmar2016NatPhy}%
  \BibitemOpen
  \bibfield  {author} {\bibinfo {author} {\bibfnamefont {F.}~\bibnamefont
  {Kretzschmar}}, \bibinfo {author} {\bibfnamefont {T.}~\bibnamefont {Bohm}},
  \bibinfo {author} {\bibfnamefont {U.}~\bibnamefont {Karahasanovic}}, \bibinfo
  {author} {\bibfnamefont {B.}~\bibnamefont {Muschler}}, \bibinfo {author}
  {\bibfnamefont {A.}~\bibnamefont {Baum}}, \bibinfo {author} {\bibfnamefont
  {D.}~\bibnamefont {Jost}}, \bibinfo {author} {\bibfnamefont {J.}~\bibnamefont
  {Schmalian}}, \bibinfo {author} {\bibfnamefont {S.}~\bibnamefont {Caprara}},
  \bibinfo {author} {\bibfnamefont {M.}~\bibnamefont {Grilli}}, \bibinfo
  {author} {\bibfnamefont {C.}~\bibnamefont {Di~Castro}}, \bibinfo {author}
  {\bibfnamefont {J.~G.}\ \bibnamefont {Analytis}}, \bibinfo {author}
  {\bibfnamefont {J.~H.}\ \bibnamefont {Chu}}, \bibinfo {author} {\bibfnamefont
  {I.~R.}\ \bibnamefont {Fisher}}, \ and\ \bibinfo {author} {\bibfnamefont
  {R.}~\bibnamefont {Hackl}},\ }\bibfield  {title} {\enquote {\bibinfo {title}
  {Critical spin fluctuations and the origin of nematic order in
  {B}a({F}e$_{1-x}${C}o$_{x}$)$_2${A}s$_2$},}\ }\href
  {http://dx.doi.org/10.1038/nphys3634} {\bibfield  {journal} {\bibinfo
  {journal} {Nature Phys.}\ }\textbf {\bibinfo {volume} {12}},\ \bibinfo
  {pages} {560} (\bibinfo {year} {2016})}\BibitemShut {NoStop}%
\bibitem [{\citenamefont {Ying}\ \emph {et~al.}(2011)\citenamefont {Ying},
  \citenamefont {Wang}, \citenamefont {Wu}, \citenamefont {Xiang},
  \citenamefont {Liu}, \citenamefont {Yan}, \citenamefont {Wang}, \citenamefont
  {Zhang}, \citenamefont {Ye}, \citenamefont {Cheng}, \citenamefont {Hu},\ and\
  \citenamefont {Chen}}]{Ying2011PRL}%
  \BibitemOpen
  \bibfield  {author} {\bibinfo {author} {\bibfnamefont {J.~J.}\ \bibnamefont
  {Ying}}, \bibinfo {author} {\bibfnamefont {X.~F.}\ \bibnamefont {Wang}},
  \bibinfo {author} {\bibfnamefont {T.}~\bibnamefont {Wu}}, \bibinfo {author}
  {\bibfnamefont {Z.~J.}\ \bibnamefont {Xiang}}, \bibinfo {author}
  {\bibfnamefont {R.~H.}\ \bibnamefont {Liu}}, \bibinfo {author} {\bibfnamefont
  {Y.~J.}\ \bibnamefont {Yan}}, \bibinfo {author} {\bibfnamefont {A.~F.}\
  \bibnamefont {Wang}}, \bibinfo {author} {\bibfnamefont {M.}~\bibnamefont
  {Zhang}}, \bibinfo {author} {\bibfnamefont {G.~J.}\ \bibnamefont {Ye}},
  \bibinfo {author} {\bibfnamefont {P.}~\bibnamefont {Cheng}}, \bibinfo
  {author} {\bibfnamefont {J.~P.}\ \bibnamefont {Hu}}, \ and\ \bibinfo {author}
  {\bibfnamefont {X.~H.}\ \bibnamefont {Chen}},\ }\bibfield  {title} {\enquote
  {\bibinfo {title} {Measurements of the anisotropic in-plane resistivity of
  underdoped {F}e{A}s-based pnictide superconductors},}\ }\href
  {http://link.aps.org/doi/10.1103/PhysRevLett.107.067001} {\bibfield
  {journal} {\bibinfo  {journal} {Phys. Rev. Lett.}\ }\textbf {\bibinfo
  {volume} {107}},\ \bibinfo {pages} {067001} (\bibinfo {year}
  {2011})}\BibitemShut {NoStop}%
\bibitem [{\citenamefont {Blomberg}\ \emph {et~al.}(2013)\citenamefont
  {Blomberg}, \citenamefont {Tanatar}, \citenamefont {Fernandes}, \citenamefont
  {Mazin}, \citenamefont {Shen}, \citenamefont {Wen}, \citenamefont {Johannes},
  \citenamefont {Schmalian},\ and\ \citenamefont
  {Prozorov}}]{blomberg2013NatCom}%
  \BibitemOpen
  \bibfield  {author} {\bibinfo {author} {\bibfnamefont {E.~C.}\ \bibnamefont
  {Blomberg}}, \bibinfo {author} {\bibfnamefont {M.~A.}\ \bibnamefont
  {Tanatar}}, \bibinfo {author} {\bibfnamefont {R.~M.}\ \bibnamefont
  {Fernandes}}, \bibinfo {author} {\bibfnamefont {I.~I.}\ \bibnamefont
  {Mazin}}, \bibinfo {author} {\bibfnamefont {B.}~\bibnamefont {Shen}},
  \bibinfo {author} {\bibfnamefont {H.~H.}\ \bibnamefont {Wen}}, \bibinfo
  {author} {\bibfnamefont {M.~D.}\ \bibnamefont {Johannes}}, \bibinfo {author}
  {\bibfnamefont {J.}~\bibnamefont {Schmalian}}, \ and\ \bibinfo {author}
  {\bibfnamefont {R}~\bibnamefont {Prozorov}},\ }\bibfield  {title} {\enquote
  {\bibinfo {title} {Sign-reversal of the in-plane resistivity anisotropy in
  hole-doped iron pnictides},}\ }\href
  {http://www.nature.com/articles/ncomms2933} {\bibfield  {journal} {\bibinfo
  {journal} {Nat. Commun.}\ }\textbf {\bibinfo {volume} {4}},\ \bibinfo {pages}
  {1914} (\bibinfo {year} {2013})}\BibitemShut {NoStop}%
\bibitem [{\citenamefont {Wu}\ \emph {et~al.}(2016)\citenamefont {Wu},
  \citenamefont {Zhang}, \citenamefont {Hu}, \citenamefont {Kung},
  \citenamefont {Lee}, \citenamefont {Mao}, \citenamefont {Dai}, \citenamefont
  {Ding}, \citenamefont {Richard},\ and\ \citenamefont
  {Blumberg}}]{wu2016arxiv}%
  \BibitemOpen
  \bibfield  {author} {\bibinfo {author} {\bibfnamefont {S.~F.}\ \bibnamefont
  {Wu}}, \bibinfo {author} {\bibfnamefont {W.~L.}\ \bibnamefont {Zhang}},
  \bibinfo {author} {\bibfnamefont {D.}~\bibnamefont {Hu}}, \bibinfo {author}
  {\bibfnamefont {H.~H}\ \bibnamefont {Kung}}, \bibinfo {author} {\bibfnamefont
  {A.}~\bibnamefont {Lee}}, \bibinfo {author} {\bibfnamefont {H.~C.}\
  \bibnamefont {Mao}}, \bibinfo {author} {\bibfnamefont {P.~C.}\ \bibnamefont
  {Dai}}, \bibinfo {author} {\bibfnamefont {H.}~\bibnamefont {Ding}}, \bibinfo
  {author} {\bibfnamefont {P.}~\bibnamefont {Richard}}, \ and\ \bibinfo
  {author} {\bibfnamefont {G.}~\bibnamefont {Blumberg}},\ }\bibfield  {title}
  {\enquote {\bibinfo {title} {Collective excitations of dynamic {F}ermi
  surface deformations in {B}a{F}e$_2$({A}s$_{0.5}${P}$_{0.5}$)$_2$},}\ }\href
  {https://arxiv.org/abs/1607.06575} {\bibfield  {journal} {\bibinfo  {journal}
  {arXiv:1607.06575}\ } (\bibinfo {year} {2016})}\BibitemShut {NoStop}%
\bibitem [{\citenamefont {Goto}\ \emph {et~al.}(2011)\citenamefont {Goto},
  \citenamefont {Kurihara}, \citenamefont {Araki}, \citenamefont {Mitsumoto},
  \citenamefont {Akatsu}, \citenamefont {Nemoto}, \citenamefont {Tatematsu},\
  and\ \citenamefont {Sato}}]{GotoJPSJ}%
  \BibitemOpen
  \bibfield  {author} {\bibinfo {author} {\bibfnamefont {T.}~\bibnamefont
  {Goto}}, \bibinfo {author} {\bibfnamefont {R.}~\bibnamefont {Kurihara}},
  \bibinfo {author} {\bibfnamefont {K.}~\bibnamefont {Araki}}, \bibinfo
  {author} {\bibfnamefont {K.}~\bibnamefont {Mitsumoto}}, \bibinfo {author}
  {\bibfnamefont {M.}~\bibnamefont {Akatsu}}, \bibinfo {author} {\bibfnamefont
  {Y.}~\bibnamefont {Nemoto}}, \bibinfo {author} {\bibfnamefont
  {S.}~\bibnamefont {Tatematsu}}, \ and\ \bibinfo {author} {\bibfnamefont
  {M.}~\bibnamefont {Sato}},\ }\bibfield  {title} {\enquote {\bibinfo {title}
  {Quadrupole effects of layered iron pnictide superconductor
  {B}a({F}e$_{0.9}${C}o$_{0.1}$)$_2${A}s$_2$},}\ }\href
  {http://dx.doi.org/10.1143/JPSJ.80.073702} {\bibfield  {journal} {\bibinfo
  {journal} {J. Phys. Soc. Jpn.}\ }\textbf {\bibinfo {volume} {80}},\ \bibinfo
  {pages} {073702} (\bibinfo {year} {2011})}\BibitemShut {NoStop}%
\bibitem [{\citenamefont {Fernandes}\ \emph {et~al.}(2010)\citenamefont
  {Fernandes}, \citenamefont {VanBebber}, \citenamefont {Bhattacharya},
  \citenamefont {Chandra}, \citenamefont {Keppens}, \citenamefont {Mandrus},
  \citenamefont {McGuire}, \citenamefont {Sales}, \citenamefont {Sefat},\ and\
  \citenamefont {Schmalian}}]{Fernandes2010PRL}%
  \BibitemOpen
  \bibfield  {author} {\bibinfo {author} {\bibfnamefont {R.~M.}\ \bibnamefont
  {Fernandes}}, \bibinfo {author} {\bibfnamefont {L.~H.}\ \bibnamefont
  {VanBebber}}, \bibinfo {author} {\bibfnamefont {S.}~\bibnamefont
  {Bhattacharya}}, \bibinfo {author} {\bibfnamefont {P.}~\bibnamefont
  {Chandra}}, \bibinfo {author} {\bibfnamefont {V.}~\bibnamefont {Keppens}},
  \bibinfo {author} {\bibfnamefont {D.}~\bibnamefont {Mandrus}}, \bibinfo
  {author} {\bibfnamefont {M.~A.}\ \bibnamefont {McGuire}}, \bibinfo {author}
  {\bibfnamefont {B.~C.}\ \bibnamefont {Sales}}, \bibinfo {author}
  {\bibfnamefont {A.~S.}\ \bibnamefont {Sefat}}, \ and\ \bibinfo {author}
  {\bibfnamefont {J.}~\bibnamefont {Schmalian}},\ }\bibfield  {title} {\enquote
  {\bibinfo {title} {Effects of nematic fluctuations on the elastic properties
  of iron arsenide superconductors},}\ }\href
  {http://link.aps.org/doi/10.1103/PhysRevLett.105.157003} {\bibfield
  {journal} {\bibinfo  {journal} {Phys. Rev. Lett.}\ }\textbf {\bibinfo
  {volume} {105}},\ \bibinfo {pages} {157003} (\bibinfo {year}
  {2010})}\BibitemShut {NoStop}%
\bibitem [{\citenamefont {Xu}\ \emph {et~al.}(2011{\natexlab{a}})\citenamefont
  {Xu}, \citenamefont {Richard}, \citenamefont {Nakayama}, \citenamefont
  {Kawahara}, \citenamefont {Sekiba}, \citenamefont {Qian}, \citenamefont
  {Neupane}, \citenamefont {Souma}, \citenamefont {Sato}, \citenamefont
  {Takahashi}, \citenamefont {Luo}, \citenamefont {Wen}, \citenamefont {Chen},
  \citenamefont {Wang}, \citenamefont {Wang}, \citenamefont {Fang},
  \citenamefont {Dai},\ and\ \citenamefont {Ding}}]{xu2011NatCom}%
  \BibitemOpen
  \bibfield  {author} {\bibinfo {author} {\bibfnamefont {Y.~M.}\ \bibnamefont
  {Xu}}, \bibinfo {author} {\bibfnamefont {P.}~\bibnamefont {Richard}},
  \bibinfo {author} {\bibfnamefont {K.}~\bibnamefont {Nakayama}}, \bibinfo
  {author} {\bibfnamefont {T.}~\bibnamefont {Kawahara}}, \bibinfo {author}
  {\bibfnamefont {Y.}~\bibnamefont {Sekiba}}, \bibinfo {author} {\bibfnamefont
  {T.}~\bibnamefont {Qian}}, \bibinfo {author} {\bibfnamefont {M.}~\bibnamefont
  {Neupane}}, \bibinfo {author} {\bibfnamefont {S.}~\bibnamefont {Souma}},
  \bibinfo {author} {\bibfnamefont {T.}~\bibnamefont {Sato}}, \bibinfo {author}
  {\bibfnamefont {T.}~\bibnamefont {Takahashi}}, \bibinfo {author}
  {\bibfnamefont {H.~Q.}\ \bibnamefont {Luo}}, \bibinfo {author} {\bibfnamefont
  {H.~H.}\ \bibnamefont {Wen}}, \bibinfo {author} {\bibfnamefont {G.~F.}\
  \bibnamefont {Chen}}, \bibinfo {author} {\bibfnamefont {N.~L.}\ \bibnamefont
  {Wang}}, \bibinfo {author} {\bibfnamefont {Z.}~\bibnamefont {Wang}}, \bibinfo
  {author} {\bibfnamefont {Z.}~\bibnamefont {Fang}}, \bibinfo {author}
  {\bibfnamefont {X.}~\bibnamefont {Dai}}, \ and\ \bibinfo {author}
  {\bibfnamefont {H.}~\bibnamefont {Ding}},\ }\bibfield  {title} {\enquote
  {\bibinfo {title} {{F}ermi surface dichotomy of the superconducting gap and
  pseudogap in underdoped pnictides},}\ }\href
  {http://dx.doi.org/10.1038/ncomms1394} {\bibfield  {journal} {\bibinfo
  {journal} {Nat. Commun.}\ }\textbf {\bibinfo {volume} {2}},\ \bibinfo {pages}
  {394} (\bibinfo {year} {2011}{\natexlab{a}})}\BibitemShut {NoStop}%
\bibitem [{\citenamefont {Gong}\ \emph {et~al.}(2015)\citenamefont {Gong},
  \citenamefont {Hou}, \citenamefont {Zhu}, \citenamefont {Jie}, \citenamefont
  {Gu}, \citenamefont {Shen}, \citenamefont {Ren}, \citenamefont {Li},\ and\
  \citenamefont {Shan}}]{Jing2015CPB}%
  \BibitemOpen
  \bibfield  {author} {\bibinfo {author} {\bibfnamefont {J.}~\bibnamefont
  {Gong}}, \bibinfo {author} {\bibfnamefont {X.~Y.}\ \bibnamefont {Hou}},
  \bibinfo {author} {\bibfnamefont {J.}~\bibnamefont {Zhu}}, \bibinfo {author}
  {\bibfnamefont {Y.~Y.}\ \bibnamefont {Jie}}, \bibinfo {author} {\bibfnamefont
  {Y.~D.}\ \bibnamefont {Gu}}, \bibinfo {author} {\bibfnamefont
  {B.}~\bibnamefont {Shen}}, \bibinfo {author} {\bibfnamefont {C.}~\bibnamefont
  {Ren}}, \bibinfo {author} {\bibfnamefont {C.~H.}\ \bibnamefont {Li}}, \ and\
  \bibinfo {author} {\bibfnamefont {L.}~\bibnamefont {Shan}},\ }\bibfield
  {title} {\enquote {\bibinfo {title} {Observation of mode-like features in
  tunneling spectra of iron-based superconductors},}\ }\href
  {http://stacks.iop.org/1674-1056/24/i=7/a=077402} {\bibfield  {journal}
  {\bibinfo  {journal} {Chin. Phys. B}\ }\textbf {\bibinfo {volume} {24}},\
  \bibinfo {pages} {077402} (\bibinfo {year} {2015})}\BibitemShut {NoStop}%
\bibitem [{\citenamefont {Castellan}\ \emph {et~al.}(2011)\citenamefont
  {Castellan}, \citenamefont {Rosenkranz}, \citenamefont {Goremychkin},
  \citenamefont {Chung}, \citenamefont {Todorov}, \citenamefont {Kanatzidis},
  \citenamefont {Eremin}, \citenamefont {Knolle}, \citenamefont {Chubukov},
  \citenamefont {Maiti}, \citenamefont {Norman}, \citenamefont {Weber},
  \citenamefont {Claus}, \citenamefont {Guidi}, \citenamefont {Bewley},\ and\
  \citenamefont {Osborn}}]{Castellan2011PRL}%
  \BibitemOpen
  \bibfield  {author} {\bibinfo {author} {\bibfnamefont {J.~P.}\ \bibnamefont
  {Castellan}}, \bibinfo {author} {\bibfnamefont {S.}~\bibnamefont
  {Rosenkranz}}, \bibinfo {author} {\bibfnamefont {E.~A.}\ \bibnamefont
  {Goremychkin}}, \bibinfo {author} {\bibfnamefont {D.~Y.}\ \bibnamefont
  {Chung}}, \bibinfo {author} {\bibfnamefont {I.~S.}\ \bibnamefont {Todorov}},
  \bibinfo {author} {\bibfnamefont {M.~G.}\ \bibnamefont {Kanatzidis}},
  \bibinfo {author} {\bibfnamefont {I.}~\bibnamefont {Eremin}}, \bibinfo
  {author} {\bibfnamefont {J.}~\bibnamefont {Knolle}}, \bibinfo {author}
  {\bibfnamefont {A.~V.}\ \bibnamefont {Chubukov}}, \bibinfo {author}
  {\bibfnamefont {S.}~\bibnamefont {Maiti}}, \bibinfo {author} {\bibfnamefont
  {M.~R.}\ \bibnamefont {Norman}}, \bibinfo {author} {\bibfnamefont
  {F.}~\bibnamefont {Weber}}, \bibinfo {author} {\bibfnamefont
  {H.}~\bibnamefont {Claus}}, \bibinfo {author} {\bibfnamefont
  {T.}~\bibnamefont {Guidi}}, \bibinfo {author} {\bibfnamefont {R.~I.}\
  \bibnamefont {Bewley}}, \ and\ \bibinfo {author} {\bibfnamefont
  {R.}~\bibnamefont {Osborn}},\ }\bibfield  {title} {\enquote {\bibinfo {title}
  {Effect of {F}ermi surface nesting on resonant spin excitations in
  {B}a$_{1-x}${K}$_{x}${F}e$_{2}${A}s$_2$},}\ }\href
  {http://link.aps.org/doi/10.1103/PhysRevLett.107.177003} {\bibfield
  {journal} {\bibinfo  {journal} {Phys. Rev. Lett.}\ }\textbf {\bibinfo
  {volume} {107}},\ \bibinfo {pages} {177003} (\bibinfo {year}
  {2011})}\BibitemShut {NoStop}%
\bibitem [{\citenamefont {Ding}\ \emph {et~al.}(2008)\citenamefont {Ding},
  \citenamefont {Richard}, \citenamefont {Nakayama}, \citenamefont {Sugawara},
  \citenamefont {Arakane}, \citenamefont {Sekiba}, \citenamefont {Takayama},
  \citenamefont {Souma}, \citenamefont {Sato}, \citenamefont {Takahashi},
  \citenamefont {Wang}, \citenamefont {Dai}, \citenamefont {Fang},
  \citenamefont {Chen}, \citenamefont {Luo},\ and\ \citenamefont
  {Wang}}]{ding2008EPLgap}%
  \BibitemOpen
  \bibfield  {author} {\bibinfo {author} {\bibfnamefont {H.}~\bibnamefont
  {Ding}}, \bibinfo {author} {\bibfnamefont {P.}~\bibnamefont {Richard}},
  \bibinfo {author} {\bibfnamefont {K.}~\bibnamefont {Nakayama}}, \bibinfo
  {author} {\bibfnamefont {K.}~\bibnamefont {Sugawara}}, \bibinfo {author}
  {\bibfnamefont {T.}~\bibnamefont {Arakane}}, \bibinfo {author} {\bibfnamefont
  {Y.}~\bibnamefont {Sekiba}}, \bibinfo {author} {\bibfnamefont
  {A.}~\bibnamefont {Takayama}}, \bibinfo {author} {\bibfnamefont
  {S.}~\bibnamefont {Souma}}, \bibinfo {author} {\bibfnamefont
  {T.}~\bibnamefont {Sato}}, \bibinfo {author} {\bibfnamefont {T.}~\bibnamefont
  {Takahashi}}, \bibinfo {author} {\bibfnamefont {Z.}~\bibnamefont {Wang}},
  \bibinfo {author} {\bibfnamefont {X.}~\bibnamefont {Dai}}, \bibinfo {author}
  {\bibfnamefont {Z.}~\bibnamefont {Fang}}, \bibinfo {author} {\bibfnamefont
  {G.~F.}\ \bibnamefont {Chen}}, \bibinfo {author} {\bibfnamefont {J.~L.}\
  \bibnamefont {Luo}}, \ and\ \bibinfo {author} {\bibfnamefont {N.~L.}\
  \bibnamefont {Wang}},\ }\bibfield  {title} {\enquote {\bibinfo {title}
  {Observation of {F}ermi-surface-dependent nodeless superconducting gaps in
  {B}a$_{0.6}${K}$_{0.4}${F}e$_{2}${A}s$_2$},}\ }\href
  {http://stacks.iop.org/0295-5075/83/i=4/a=47001} {\bibfield  {journal}
  {\bibinfo  {journal} {Europhys. Lett.}\ }\textbf {\bibinfo {volume} {83}},\
  \bibinfo {pages} {47001} (\bibinfo {year} {2008})}\BibitemShut {NoStop}%
\bibitem [{\citenamefont {Zhao}\ \emph {et~al.}(2008)\citenamefont {Zhao},
  \citenamefont {Liu}, \citenamefont {Zhang}, \citenamefont {Meng},
  \citenamefont {Jia}, \citenamefont {Liu}, \citenamefont {Dong}, \citenamefont
  {Chen}, \citenamefont {Luo}, \citenamefont {Wang}, \citenamefont {Lu},
  \citenamefont {Wang}, \citenamefont {Zhou}, \citenamefont {Zhu},
  \citenamefont {Wang}, \citenamefont {Xu}, \citenamefont {Chen},\ and\
  \citenamefont {Zhou}}]{L_Zhao}%
  \BibitemOpen
  \bibfield  {author} {\bibinfo {author} {\bibfnamefont {L.}~\bibnamefont
  {Zhao}}, \bibinfo {author} {\bibfnamefont {H.~Y.}\ \bibnamefont {Liu}},
  \bibinfo {author} {\bibfnamefont {W.~T.}\ \bibnamefont {Zhang}}, \bibinfo
  {author} {\bibfnamefont {J.~Q.}\ \bibnamefont {Meng}}, \bibinfo {author}
  {\bibfnamefont {X.~W.}\ \bibnamefont {Jia}}, \bibinfo {author} {\bibfnamefont
  {G.~D.}\ \bibnamefont {Liu}}, \bibinfo {author} {\bibfnamefont {X.~L}\
  \bibnamefont {Dong}}, \bibinfo {author} {\bibfnamefont {G.~F.}\ \bibnamefont
  {Chen}}, \bibinfo {author} {\bibfnamefont {J.~L.}\ \bibnamefont {Luo}},
  \bibinfo {author} {\bibfnamefont {N.~L.}\ \bibnamefont {Wang}}, \bibinfo
  {author} {\bibfnamefont {W.}~\bibnamefont {Lu}}, \bibinfo {author}
  {\bibfnamefont {G.~L.}\ \bibnamefont {Wang}}, \bibinfo {author}
  {\bibfnamefont {Y.}~\bibnamefont {Zhou}}, \bibinfo {author} {\bibfnamefont
  {Y.}~\bibnamefont {Zhu}}, \bibinfo {author} {\bibfnamefont {X.~Y.}\
  \bibnamefont {Wang}}, \bibinfo {author} {\bibfnamefont {Z.~Y}\ \bibnamefont
  {Xu}}, \bibinfo {author} {\bibfnamefont {C.~T.}\ \bibnamefont {Chen}}, \ and\
  \bibinfo {author} {\bibfnamefont {X.~J.}\ \bibnamefont {Zhou}},\ }\bibfield
  {title} {\enquote {\bibinfo {title} {Multiple nodeless superconducting gaps
  in {B}a$_{0.6}${K}$_{0.4}${F}e$_{2}${A}s$_2$ superconductor from
  angle-resolved photoemission spectroscopy},}\ }\href
  {http://stacks.iop.org/0256-307X/25/i=12/a=061} {\bibfield  {journal}
  {\bibinfo  {journal} {Chin. Phys. Lett.}\ }\textbf {\bibinfo {volume} {25}},\
  \bibinfo {pages} {4402} (\bibinfo {year} {2008})}\BibitemShut {NoStop}%
\bibitem [{\citenamefont {Nakayama}\ \emph {et~al.}(2009)\citenamefont
  {Nakayama}, \citenamefont {Sato}, \citenamefont {Richard}, \citenamefont
  {Xu}, \citenamefont {Sekiba}, \citenamefont {Souma}, \citenamefont {Chen},
  \citenamefont {Luo}, \citenamefont {Wang}, \citenamefont {Ding},\ and\
  \citenamefont {Takahashi}}]{nakayama2009EPLgap}%
  \BibitemOpen
  \bibfield  {author} {\bibinfo {author} {\bibfnamefont {K.}~\bibnamefont
  {Nakayama}}, \bibinfo {author} {\bibfnamefont {T.}~\bibnamefont {Sato}},
  \bibinfo {author} {\bibfnamefont {P.}~\bibnamefont {Richard}}, \bibinfo
  {author} {\bibfnamefont {Y.~M.}\ \bibnamefont {Xu}}, \bibinfo {author}
  {\bibfnamefont {Y.}~\bibnamefont {Sekiba}}, \bibinfo {author} {\bibfnamefont
  {S.}~\bibnamefont {Souma}}, \bibinfo {author} {\bibfnamefont {G.~F.}\
  \bibnamefont {Chen}}, \bibinfo {author} {\bibfnamefont {J.~L.}\ \bibnamefont
  {Luo}}, \bibinfo {author} {\bibfnamefont {N.~L.}\ \bibnamefont {Wang}},
  \bibinfo {author} {\bibfnamefont {H.}~\bibnamefont {Ding}}, \ and\ \bibinfo
  {author} {\bibfnamefont {T.}~\bibnamefont {Takahashi}},\ }\bibfield  {title}
  {\enquote {\bibinfo {title} {Superconducting gap symmetry of
  {B}a$_{0.6}${K}$_{0.4}${F}e$_2${A}s$_2$ studied by angle-resolved
  photoemission spectroscopy},}\ }\href
  {http://stacks.iop.org/0295-5075/85/i=6/a=67002} {\bibfield  {journal}
  {\bibinfo  {journal} {Europhys. Lett.}\ }\textbf {\bibinfo {volume} {85}},\
  \bibinfo {pages} {67002} (\bibinfo {year} {2009})}\BibitemShut {NoStop}%
\bibitem [{\citenamefont {Yin}\ \emph {et~al.}(2016)\citenamefont {Yin},
  \citenamefont {Li}, \citenamefont {Wu}, \citenamefont {Li}, \citenamefont
  {Wu}, \citenamefont {Wang}, \citenamefont {Ting}, \citenamefont {Hor},
  \citenamefont {Liang}, \citenamefont {Zhang}, \citenamefont {Dai},
  \citenamefont {Wang}, \citenamefont {Jin}, \citenamefont {Chen},
  \citenamefont {Hu}, \citenamefont {Wang},\ and\ \citenamefont
  {Pan}}]{YinJX2016arxiv}%
  \BibitemOpen
  \bibfield  {author} {\bibinfo {author} {\bibfnamefont {J.~X.}\ \bibnamefont
  {Yin}}, \bibinfo {author} {\bibfnamefont {A.}~\bibnamefont {Li}}, \bibinfo
  {author} {\bibfnamefont {X.~X.}\ \bibnamefont {Wu}}, \bibinfo {author}
  {\bibfnamefont {J.}~\bibnamefont {Li}}, \bibinfo {author} {\bibfnamefont
  {Z.}~\bibnamefont {Wu}}, \bibinfo {author} {\bibfnamefont {J.~H.}\
  \bibnamefont {Wang}}, \bibinfo {author} {\bibfnamefont {C.~S.}\ \bibnamefont
  {Ting}}, \bibinfo {author} {\bibfnamefont {P.~H.}\ \bibnamefont {Hor}},
  \bibinfo {author} {\bibfnamefont {X.~J.}\ \bibnamefont {Liang}}, \bibinfo
  {author} {\bibfnamefont {C.~L.}\ \bibnamefont {Zhang}}, \bibinfo {author}
  {\bibfnamefont {P.~C.}\ \bibnamefont {Dai}}, \bibinfo {author} {\bibfnamefont
  {X.~C.}\ \bibnamefont {Wang}}, \bibinfo {author} {\bibfnamefont {C.~Q.}\
  \bibnamefont {Jin}}, \bibinfo {author} {\bibfnamefont {G.~F.}\ \bibnamefont
  {Chen}}, \bibinfo {author} {\bibfnamefont {J.~P.}\ \bibnamefont {Hu}},
  \bibinfo {author} {\bibfnamefont {Z.~Q.}\ \bibnamefont {Wang}}, \ and\
  \bibinfo {author} {\bibfnamefont {S.~H.}\ \bibnamefont {Pan}},\ }\bibfield
  {title} {\enquote {\bibinfo {title} {Real-space orbital-selective probing of
  the cooper pairing in iron pnictides},}\ }\href
  {https://arxiv.org/abs/1602.04949} {\bibfield  {journal} {\bibinfo  {journal}
  {arXiv:1602.04949}\ } (\bibinfo {year} {2016})}\BibitemShut {NoStop}%
\bibitem [{\citenamefont {Nakayama}\ \emph {et~al.}(2011)\citenamefont
  {Nakayama}, \citenamefont {Sato}, \citenamefont {Richard}, \citenamefont
  {Xu}, \citenamefont {Kawahara}, \citenamefont {Umezawa}, \citenamefont
  {Qian}, \citenamefont {Neupane}, \citenamefont {Chen}, \citenamefont {Ding},\
  and\ \citenamefont {Takahashi}}]{Nakayama_PRB2011}%
  \BibitemOpen
  \bibfield  {author} {\bibinfo {author} {\bibfnamefont {K.}~\bibnamefont
  {Nakayama}}, \bibinfo {author} {\bibfnamefont {T.}~\bibnamefont {Sato}},
  \bibinfo {author} {\bibfnamefont {P.}~\bibnamefont {Richard}}, \bibinfo
  {author} {\bibfnamefont {Y.~M.}\ \bibnamefont {Xu}}, \bibinfo {author}
  {\bibfnamefont {T.}~\bibnamefont {Kawahara}}, \bibinfo {author}
  {\bibfnamefont {K.}~\bibnamefont {Umezawa}}, \bibinfo {author} {\bibfnamefont
  {T.}~\bibnamefont {Qian}}, \bibinfo {author} {\bibfnamefont {M.}~\bibnamefont
  {Neupane}}, \bibinfo {author} {\bibfnamefont {G.~F.}\ \bibnamefont {Chen}},
  \bibinfo {author} {\bibfnamefont {H.}~\bibnamefont {Ding}}, \ and\ \bibinfo
  {author} {\bibfnamefont {T.}~\bibnamefont {Takahashi}},\ }\bibfield  {title}
  {\enquote {\bibinfo {title} {Universality of superconducting gaps in
  overdoped {B}a$_{0.3}${K}$_{0.7}${F}e$_{2}${A}s$_2$ observed by
  angle-resolved photoemission spectroscopy},}\ }\href
  {http://link.aps.org/doi/10.1103/PhysRevB.83.020501} {\bibfield  {journal}
  {\bibinfo  {journal} {Phys. Rev. B}\ }\textbf {\bibinfo {volume} {83}},\
  \bibinfo {pages} {020501} (\bibinfo {year} {2011})}\BibitemShut {NoStop}%
\bibitem [{\citenamefont {Lee}\ \emph {et~al.}(2016)\citenamefont {Lee},
  \citenamefont {Kihou}, \citenamefont {Park}, \citenamefont {Horigane},
  \citenamefont {Fujita}, \citenamefont {Wa{\ss}er}, \citenamefont {Qureshi},
  \citenamefont {Sidis}, \citenamefont {Akimitsu},\ and\ \citenamefont
  {Braden}}]{Lee2016SciRep}%
  \BibitemOpen
  \bibfield  {author} {\bibinfo {author} {\bibfnamefont {C.~H.}\ \bibnamefont
  {Lee}}, \bibinfo {author} {\bibfnamefont {K.}~\bibnamefont {Kihou}}, \bibinfo
  {author} {\bibfnamefont {J.~T.}\ \bibnamefont {Park}}, \bibinfo {author}
  {\bibfnamefont {K.}~\bibnamefont {Horigane}}, \bibinfo {author}
  {\bibfnamefont {K.}~\bibnamefont {Fujita}}, \bibinfo {author} {\bibfnamefont
  {F.}~\bibnamefont {Wa{\ss}er}}, \bibinfo {author} {\bibfnamefont
  {N.}~\bibnamefont {Qureshi}}, \bibinfo {author} {\bibfnamefont
  {Y.}~\bibnamefont {Sidis}}, \bibinfo {author} {\bibfnamefont
  {J.}~\bibnamefont {Akimitsu}}, \ and\ \bibinfo {author} {\bibfnamefont
  {M.}~\bibnamefont {Braden}},\ }\bibfield  {title} {\enquote {\bibinfo {title}
  {Suppression of spin-exciton state in hole overdoped iron-based
  superconductors},}\ }\href {http://dx.doi.org/10.1038/srep23424} {\bibfield
  {journal} {\bibinfo  {journal} {Sci. Rep.}\ }\textbf {\bibinfo {volume}
  {6}},\ \bibinfo {pages} {23424} (\bibinfo {year} {2016})}\BibitemShut
  {NoStop}%
\bibitem [{\citenamefont {Xu}\ \emph {et~al.}(2011{\natexlab{b}})\citenamefont
  {Xu}, \citenamefont {Huang}, \citenamefont {Cui}, \citenamefont {Razzoli},
  \citenamefont {Radovic}, \citenamefont {Shi}, \citenamefont {Chen},
  \citenamefont {Zheng}, \citenamefont {Wang}, \citenamefont {Zhang},
  \citenamefont {Dai}, \citenamefont {Hu}, \citenamefont {Wang},\ and\
  \citenamefont {Ding}}]{xu2011NatPhysics}%
  \BibitemOpen
  \bibfield  {author} {\bibinfo {author} {\bibfnamefont {Y.~M.}\ \bibnamefont
  {Xu}}, \bibinfo {author} {\bibfnamefont {Y.~B.}\ \bibnamefont {Huang}},
  \bibinfo {author} {\bibfnamefont {X.~Y.}\ \bibnamefont {Cui}}, \bibinfo
  {author} {\bibfnamefont {E.}~\bibnamefont {Razzoli}}, \bibinfo {author}
  {\bibfnamefont {M.}~\bibnamefont {Radovic}}, \bibinfo {author} {\bibfnamefont
  {M.}~\bibnamefont {Shi}}, \bibinfo {author} {\bibfnamefont {G.~F.}\
  \bibnamefont {Chen}}, \bibinfo {author} {\bibfnamefont {P.}~\bibnamefont
  {Zheng}}, \bibinfo {author} {\bibfnamefont {N.~L.}\ \bibnamefont {Wang}},
  \bibinfo {author} {\bibfnamefont {C.~L.}\ \bibnamefont {Zhang}}, \bibinfo
  {author} {\bibfnamefont {P.~C.}\ \bibnamefont {Dai}}, \bibinfo {author}
  {\bibfnamefont {J.~P.}\ \bibnamefont {Hu}}, \bibinfo {author} {\bibfnamefont
  {Z.}~\bibnamefont {Wang}}, \ and\ \bibinfo {author} {\bibfnamefont
  {H.}~\bibnamefont {Ding}},\ }\bibfield  {title} {\enquote {\bibinfo {title}
  {Observation of a ubiquitous three-dimensional superconducting gap function
  in optimally doped {B}a$_{0.6}${K}$_{0.4}${F}e$_{2}${A}s$_2$},}\ }\href
  {\doibase 10.1038/nphys1879} {\bibfield  {journal} {\bibinfo  {journal}
  {Nature Phys.}\ }\textbf {\bibinfo {volume} {7}},\ \bibinfo {pages} {198}
  (\bibinfo {year} {2011}{\natexlab{b}})}\BibitemShut {NoStop}%
\bibitem [{\citenamefont {Li}\ \emph {et~al.}(2008)\citenamefont {Li},
  \citenamefont {Hu}, \citenamefont {Dong}, \citenamefont {Li}, \citenamefont
  {Zheng}, \citenamefont {Chen}, \citenamefont {Luo},\ and\ \citenamefont
  {Wang}}]{Li2008PRL}%
  \BibitemOpen
  \bibfield  {author} {\bibinfo {author} {\bibfnamefont {G.}~\bibnamefont
  {Li}}, \bibinfo {author} {\bibfnamefont {W.~Z.}\ \bibnamefont {Hu}}, \bibinfo
  {author} {\bibfnamefont {J.}~\bibnamefont {Dong}}, \bibinfo {author}
  {\bibfnamefont {Z.}~\bibnamefont {Li}}, \bibinfo {author} {\bibfnamefont
  {P.}~\bibnamefont {Zheng}}, \bibinfo {author} {\bibfnamefont {G.~F.}\
  \bibnamefont {Chen}}, \bibinfo {author} {\bibfnamefont {J.~L.}\ \bibnamefont
  {Luo}}, \ and\ \bibinfo {author} {\bibfnamefont {N.~L.}\ \bibnamefont
  {Wang}},\ }\bibfield  {title} {\enquote {\bibinfo {title} {Probing the
  superconducting energy gap from infrared spectroscopy on a
  {B}a$_{0.6}${K}$_{0.4}${F}e$_{2}${A}s$_2$ single crystal with
  ${T}_{c}=37\text{ }\text{ }\mathrm{K}$},}\ }\href
  {http://link.aps.org/doi/10.1103/PhysRevLett.101.107004} {\bibfield
  {journal} {\bibinfo  {journal} {Phys. Rev. Lett.}\ }\textbf {\bibinfo
  {volume} {101}},\ \bibinfo {pages} {107004} (\bibinfo {year}
  {2008})}\BibitemShut {NoStop}%
\bibitem [{\citenamefont {Luo}\ \emph {et~al.}(2009)\citenamefont {Luo},
  \citenamefont {Tanatar}, \citenamefont {Reid}, \citenamefont {Shakeripour},
  \citenamefont {Doiron-Leyraud}, \citenamefont {Ni}, \citenamefont {Bud'ko},
  \citenamefont {Canfield}, \citenamefont {Luo}, \citenamefont {Wang},
  \citenamefont {Wen}, \citenamefont {Prozorov},\ and\ \citenamefont
  {Taillefer}}]{XG_LuoPRB2009}%
  \BibitemOpen
  \bibfield  {author} {\bibinfo {author} {\bibfnamefont {X.~G.}\ \bibnamefont
  {Luo}}, \bibinfo {author} {\bibfnamefont {M.~A.}\ \bibnamefont {Tanatar}},
  \bibinfo {author} {\bibfnamefont {J.~P.}\ \bibnamefont {Reid}}, \bibinfo
  {author} {\bibfnamefont {H.}~\bibnamefont {Shakeripour}}, \bibinfo {author}
  {\bibfnamefont {N.}~\bibnamefont {Doiron-Leyraud}}, \bibinfo {author}
  {\bibfnamefont {N.}~\bibnamefont {Ni}}, \bibinfo {author} {\bibfnamefont
  {S.~L.}\ \bibnamefont {Bud'ko}}, \bibinfo {author} {\bibfnamefont {P.~C.}\
  \bibnamefont {Canfield}}, \bibinfo {author} {\bibfnamefont {H.~Q}\
  \bibnamefont {Luo}}, \bibinfo {author} {\bibfnamefont {Z.~S}\ \bibnamefont
  {Wang}}, \bibinfo {author} {\bibfnamefont {H.~H}\ \bibnamefont {Wen}},
  \bibinfo {author} {\bibfnamefont {R.}~\bibnamefont {Prozorov}}, \ and\
  \bibinfo {author} {\bibfnamefont {L.}~\bibnamefont {Taillefer}},\ }\bibfield
  {title} {\enquote {\bibinfo {title} {Quasiparticle heat transport in
  single-crystalline {B}a$_{1-x}${K}$_{x}${F}e$_{2}${A}s$_2$ : Evidence for a
  $k$-dependent superconducting gap without nodes},}\ }\href
  {http://link.aps.org/doi/10.1103/PhysRevB.80.140503} {\bibfield  {journal}
  {\bibinfo  {journal} {Phys. Rev. B}\ }\textbf {\bibinfo {volume} {80}},\
  \bibinfo {pages} {140503} (\bibinfo {year} {2009})}\BibitemShut {NoStop}%
\bibitem [{\citenamefont {B{\"o}hmer}\ \emph {et~al.}(2015)\citenamefont
  {B{\"o}hmer}, \citenamefont {Hardy}, \citenamefont {Wang}, \citenamefont
  {Wolf}, \citenamefont {Schweiss},\ and\ \citenamefont
  {Meingast}}]{bohmer2015NatCom}%
  \BibitemOpen
  \bibfield  {author} {\bibinfo {author} {\bibfnamefont {A.~E.}\ \bibnamefont
  {B{\"o}hmer}}, \bibinfo {author} {\bibfnamefont {F.}~\bibnamefont {Hardy}},
  \bibinfo {author} {\bibfnamefont {L.}~\bibnamefont {Wang}}, \bibinfo {author}
  {\bibfnamefont {T.}~\bibnamefont {Wolf}}, \bibinfo {author} {\bibfnamefont
  {P.}~\bibnamefont {Schweiss}}, \ and\ \bibinfo {author} {\bibfnamefont
  {C.}~\bibnamefont {Meingast}},\ }\bibfield  {title} {\enquote {\bibinfo
  {title} {Superconductivity-induced re-entrance of the orthorhombic distortion
  in {B}a$_{1-x}${K}$_{x}${F}e$_{2}${A}s$_2$},}\ }\href
  {http://www.nature.com/articles/ncomms8911} {\bibfield  {journal} {\bibinfo
  {journal} {Nat. Commun.}\ }\textbf {\bibinfo {volume} {6}} (\bibinfo {year}
  {2015})}\BibitemShut {NoStop}%
\bibitem [{\citenamefont {Allred}\ \emph {et~al.}(2015)\citenamefont {Allred},
  \citenamefont {Avci}, \citenamefont {Chung}, \citenamefont {Claus},
  \citenamefont {Khalyavin}, \citenamefont {Manuel}, \citenamefont {Taddei},
  \citenamefont {Kanatzidis}, \citenamefont {Rosenkranz}, \citenamefont
  {Osborn},\ and\ \citenamefont {Chmaissem}}]{Allred_PRB92}%
  \BibitemOpen
  \bibfield  {author} {\bibinfo {author} {\bibfnamefont {J.~M.}\ \bibnamefont
  {Allred}}, \bibinfo {author} {\bibfnamefont {S.}~\bibnamefont {Avci}},
  \bibinfo {author} {\bibfnamefont {D.~Y.}\ \bibnamefont {Chung}}, \bibinfo
  {author} {\bibfnamefont {H.}~\bibnamefont {Claus}}, \bibinfo {author}
  {\bibfnamefont {D.~D.}\ \bibnamefont {Khalyavin}}, \bibinfo {author}
  {\bibfnamefont {P.}~\bibnamefont {Manuel}}, \bibinfo {author} {\bibfnamefont
  {K.~M.}\ \bibnamefont {Taddei}}, \bibinfo {author} {\bibfnamefont {M.~G.}\
  \bibnamefont {Kanatzidis}}, \bibinfo {author} {\bibfnamefont
  {S.}~\bibnamefont {Rosenkranz}}, \bibinfo {author} {\bibfnamefont
  {R.}~\bibnamefont {Osborn}}, \ and\ \bibinfo {author} {\bibfnamefont
  {O.}~\bibnamefont {Chmaissem}},\ }\bibfield  {title} {\enquote {\bibinfo
  {title} {Tetragonal magnetic phase in {B}a$_{1-x}${K}$_{x}${F}e$_{2}${A}s$_2$
  from x-ray and neutron diffraction},}\ }\href
  {http://link.aps.org/doi/10.1103/PhysRevB.92.094515} {\bibfield  {journal}
  {\bibinfo  {journal} {Phys. Rev. B}\ }\textbf {\bibinfo {volume} {92}},\
  \bibinfo {pages} {094515} (\bibinfo {year} {2015})}\BibitemShut {NoStop}%
\bibitem [{\citenamefont {Khalyavin}\ \emph {et~al.}(2014)\citenamefont
  {Khalyavin}, \citenamefont {Lovesey}, \citenamefont {Manuel}, \citenamefont
  {Kr\"uger}, \citenamefont {Rosenkranz}, \citenamefont {Allred}, \citenamefont
  {Chmaissem},\ and\ \citenamefont {Osborn}}]{Khalyavin_PRB90}%
  \BibitemOpen
  \bibfield  {author} {\bibinfo {author} {\bibfnamefont {D.~D.}\ \bibnamefont
  {Khalyavin}}, \bibinfo {author} {\bibfnamefont {S.~W.}\ \bibnamefont
  {Lovesey}}, \bibinfo {author} {\bibfnamefont {P.}~\bibnamefont {Manuel}},
  \bibinfo {author} {\bibfnamefont {F.}~\bibnamefont {Kr\"uger}}, \bibinfo
  {author} {\bibfnamefont {S.}~\bibnamefont {Rosenkranz}}, \bibinfo {author}
  {\bibfnamefont {J.~M.}\ \bibnamefont {Allred}}, \bibinfo {author}
  {\bibfnamefont {O.}~\bibnamefont {Chmaissem}}, \ and\ \bibinfo {author}
  {\bibfnamefont {R.}~\bibnamefont {Osborn}},\ }\bibfield  {title} {\enquote
  {\bibinfo {title} {Symmetry of reentrant tetragonal phase in
  {B}a$_{1-x}${N}a$_{x}${F}e$_{2}${A}s$_2$: Magnetic versus orbital ordering
  mechanism},}\ }\href {\doibase 10.1103/PhysRevB.90.174511} {\bibfield
  {journal} {\bibinfo  {journal} {Phys. Rev. B}\ }\textbf {\bibinfo {volume}
  {90}},\ \bibinfo {pages} {174511} (\bibinfo {year} {2014})}\BibitemShut
  {NoStop}%
\bibitem [{\citenamefont {Wang}\ \emph {et~al.}(2016)\citenamefont {Wang},
  \citenamefont {Hardy}, \citenamefont {B\"ohmer}, \citenamefont {Wolf},
  \citenamefont {Schweiss},\ and\ \citenamefont {Meingast}}]{L_Wang_PRB93}%
  \BibitemOpen
  \bibfield  {author} {\bibinfo {author} {\bibfnamefont {L.}~\bibnamefont
  {Wang}}, \bibinfo {author} {\bibfnamefont {F.}~\bibnamefont {Hardy}},
  \bibinfo {author} {\bibfnamefont {A.~E.}\ \bibnamefont {B\"ohmer}}, \bibinfo
  {author} {\bibfnamefont {T.}~\bibnamefont {Wolf}}, \bibinfo {author}
  {\bibfnamefont {P.}~\bibnamefont {Schweiss}}, \ and\ \bibinfo {author}
  {\bibfnamefont {C.}~\bibnamefont {Meingast}},\ }\bibfield  {title} {\enquote
  {\bibinfo {title} {Complex phase diagram of
  {B}a$_{1-x}${N}a$_{x}${F}e$_{2}${A}s$_2$: A multitude of phases striving for
  the electronic entropy},}\ }\href {\doibase 10.1103/PhysRevB.93.014514}
  {\bibfield  {journal} {\bibinfo  {journal} {Phys. Rev. B}\ }\textbf {\bibinfo
  {volume} {93}},\ \bibinfo {pages} {014514} (\bibinfo {year}
  {2016})}\BibitemShut {NoStop}%
\bibitem [{\citenamefont {Zhang}\ \emph
  {et~al.}(2014{\natexlab{b}})\citenamefont {Zhang}, \citenamefont {Richard},
  \citenamefont {Qian}, \citenamefont {Shi}, \citenamefont {Ma}, \citenamefont
  {Zeng}, \citenamefont {Wang}, \citenamefont {Rienks}, \citenamefont {Zhang},
  \citenamefont {Dai}, \citenamefont {You}, \citenamefont {Weng}, \citenamefont
  {Wu}, \citenamefont {Hu},\ and\ \citenamefont {Ding}}]{Zhang2014PRXimpurity}%
  \BibitemOpen
  \bibfield  {author} {\bibinfo {author} {\bibfnamefont {P.}~\bibnamefont
  {Zhang}}, \bibinfo {author} {\bibfnamefont {P.}~\bibnamefont {Richard}},
  \bibinfo {author} {\bibfnamefont {T.}~\bibnamefont {Qian}}, \bibinfo {author}
  {\bibfnamefont {X.}~\bibnamefont {Shi}}, \bibinfo {author} {\bibfnamefont
  {J.}~\bibnamefont {Ma}}, \bibinfo {author} {\bibfnamefont {L.~K.}\
  \bibnamefont {Zeng}}, \bibinfo {author} {\bibfnamefont {X.~P.}\ \bibnamefont
  {Wang}}, \bibinfo {author} {\bibfnamefont {E.}~\bibnamefont {Rienks}},
  \bibinfo {author} {\bibfnamefont {C.~L.}\ \bibnamefont {Zhang}}, \bibinfo
  {author} {\bibfnamefont {P.~C}\ \bibnamefont {Dai}}, \bibinfo {author}
  {\bibfnamefont {Y.~Z.}\ \bibnamefont {You}}, \bibinfo {author} {\bibfnamefont
  {Z.~Y.}\ \bibnamefont {Weng}}, \bibinfo {author} {\bibfnamefont {X.~X.}\
  \bibnamefont {Wu}}, \bibinfo {author} {\bibfnamefont {J.~P.}\ \bibnamefont
  {Hu}}, \ and\ \bibinfo {author} {\bibfnamefont {H.}~\bibnamefont {Ding}},\
  }\bibfield  {title} {\enquote {\bibinfo {title} {Observation of
  momentum-confined in-gap impurity state in
  {B}a$_{0.6}${K}$_{0.4}${F}e$_{2}${A}s$_{2}$: Evidence for antiphase
  ${s}_{\ifmmode\pm\else\textpm\fi{}}$ pairing},}\ }\href
  {http://link.aps.org/doi/10.1103/PhysRevX.4.031001} {\bibfield  {journal}
  {\bibinfo  {journal} {Phys. Rev. X}\ }\textbf {\bibinfo {volume} {4}},\
  \bibinfo {pages} {031001} (\bibinfo {year} {2014}{\natexlab{b}})}\BibitemShut
  {NoStop}%
\bibitem [{\citenamefont {Shibauchi}\ \emph {et~al.}(2014)\citenamefont
  {Shibauchi}, \citenamefont {Carrington},\ and\ \citenamefont
  {Matsuda}}]{Shibauchi2014QCP}%
  \BibitemOpen
  \bibfield  {author} {\bibinfo {author} {\bibfnamefont {T.}~\bibnamefont
  {Shibauchi}}, \bibinfo {author} {\bibfnamefont {A.}~\bibnamefont
  {Carrington}}, \ and\ \bibinfo {author} {\bibfnamefont {Y.}~\bibnamefont
  {Matsuda}},\ }\bibfield  {title} {\enquote {\bibinfo {title} {A quantum
  critical point lying beneath the superconducting dome in iron pnictides},}\
  }\href {\doibase 10.1146/annurev-conmatphys-031113-133921} {\bibfield
  {journal} {\bibinfo  {journal} {Annu. Rev. Condens. Matter Phys.}\ }\textbf
  {\bibinfo {volume} {5}},\ \bibinfo {pages} {113} (\bibinfo {year}
  {2014})}\BibitemShut {NoStop}%
\bibitem [{\citenamefont {Fukazawa}\ \emph {et~al.}(2009)\citenamefont
  {Fukazawa}, \citenamefont {Yamada}, \citenamefont {Kondo}, \citenamefont
  {Saito}, \citenamefont {Kohori}, \citenamefont {Kuga}, \citenamefont
  {Matsumoto}, \citenamefont {Nakatsuji}, \citenamefont {Kito}, \citenamefont
  {M.~Shirage}, \citenamefont {Kihou}, \citenamefont {Takeshita}, \citenamefont
  {Lee}, \citenamefont {Iyo},\ and\ \citenamefont {Eisaki}}]{Fukazawa2009JPSJ}%
  \BibitemOpen
  \bibfield  {author} {\bibinfo {author} {\bibfnamefont {H.}~\bibnamefont
  {Fukazawa}}, \bibinfo {author} {\bibfnamefont {Y.}~\bibnamefont {Yamada}},
  \bibinfo {author} {\bibfnamefont {K.}~\bibnamefont {Kondo}}, \bibinfo
  {author} {\bibfnamefont {T.}~\bibnamefont {Saito}}, \bibinfo {author}
  {\bibfnamefont {Y.}~\bibnamefont {Kohori}}, \bibinfo {author} {\bibfnamefont
  {K.}~\bibnamefont {Kuga}}, \bibinfo {author} {\bibfnamefont {Y.}~\bibnamefont
  {Matsumoto}}, \bibinfo {author} {\bibfnamefont {S.}~\bibnamefont
  {Nakatsuji}}, \bibinfo {author} {\bibfnamefont {H.}~\bibnamefont {Kito}},
  \bibinfo {author} {\bibfnamefont {P.}~\bibnamefont {M.~Shirage}}, \bibinfo
  {author} {\bibfnamefont {K.}~\bibnamefont {Kihou}}, \bibinfo {author}
  {\bibfnamefont {N.}~\bibnamefont {Takeshita}}, \bibinfo {author}
  {\bibfnamefont {C.~H}\ \bibnamefont {Lee}}, \bibinfo {author} {\bibfnamefont
  {A}~\bibnamefont {Iyo}}, \ and\ \bibinfo {author} {\bibfnamefont
  {H.}~\bibnamefont {Eisaki}},\ }\bibfield  {title} {\enquote {\bibinfo {title}
  {Possible multiple gap superconductivity with line nodes in heavily
  hole-doped superconductor {K}{F}e$_2${A}s$_2$ studied by $^{75}${A}s nuclear
  quadrupole resonance and specific heat},}\ }\href
  {http://dx.doi.org/10.1143/JPSJ.78.083712} {\bibfield  {journal} {\bibinfo
  {journal} {J. Phys. Soc. Jpn.}\ }\textbf {\bibinfo {volume} {78}},\ \bibinfo
  {pages} {083712} (\bibinfo {year} {2009})}\BibitemShut {NoStop}%
\bibitem [{\citenamefont {Hashimoto}\ \emph {et~al.}(2010)\citenamefont
  {Hashimoto}, \citenamefont {Serafin}, \citenamefont {Tonegawa}, \citenamefont
  {Katsumata}, \citenamefont {Okazaki}, \citenamefont {Saito}, \citenamefont
  {Fukazawa}, \citenamefont {Kohori}, \citenamefont {Kihou}, \citenamefont
  {Lee}, \citenamefont {Iyo}, \citenamefont {Eisaki}, \citenamefont {Ikeda},
  \citenamefont {Matsuda}, \citenamefont {Carrington},\ and\ \citenamefont
  {Shibauchi}}]{Hashimoto2010PRB}%
  \BibitemOpen
  \bibfield  {author} {\bibinfo {author} {\bibfnamefont {K.}~\bibnamefont
  {Hashimoto}}, \bibinfo {author} {\bibfnamefont {A.}~\bibnamefont {Serafin}},
  \bibinfo {author} {\bibfnamefont {S.}~\bibnamefont {Tonegawa}}, \bibinfo
  {author} {\bibfnamefont {R.}~\bibnamefont {Katsumata}}, \bibinfo {author}
  {\bibfnamefont {R.}~\bibnamefont {Okazaki}}, \bibinfo {author} {\bibfnamefont
  {T.}~\bibnamefont {Saito}}, \bibinfo {author} {\bibfnamefont
  {H.}~\bibnamefont {Fukazawa}}, \bibinfo {author} {\bibfnamefont
  {Y.}~\bibnamefont {Kohori}}, \bibinfo {author} {\bibfnamefont
  {K.}~\bibnamefont {Kihou}}, \bibinfo {author} {\bibfnamefont {C.~H.}\
  \bibnamefont {Lee}}, \bibinfo {author} {\bibfnamefont {A.}~\bibnamefont
  {Iyo}}, \bibinfo {author} {\bibfnamefont {H.}~\bibnamefont {Eisaki}},
  \bibinfo {author} {\bibfnamefont {H.}~\bibnamefont {Ikeda}}, \bibinfo
  {author} {\bibfnamefont {Y.}~\bibnamefont {Matsuda}}, \bibinfo {author}
  {\bibfnamefont {A.}~\bibnamefont {Carrington}}, \ and\ \bibinfo {author}
  {\bibfnamefont {T.}~\bibnamefont {Shibauchi}},\ }\bibfield  {title} {\enquote
  {\bibinfo {title} {Evidence for superconducting gap nodes in the
  zone-centered hole bands of {K}{F}e$_2${A}s$_2$ from magnetic
  penetration-depth measurements},}\ }\href
  {http://link.aps.org/doi/10.1103/PhysRevB.82.014526} {\bibfield  {journal}
  {\bibinfo  {journal} {Phys. Rev. B}\ }\textbf {\bibinfo {volume} {82}},\
  \bibinfo {pages} {014526} (\bibinfo {year} {2010})}\BibitemShut {NoStop}%
\bibitem [{\citenamefont {Reid}\ \emph {et~al.}(2012)\citenamefont {Reid},
  \citenamefont {Tanatar}, \citenamefont {Juneau-Fecteau}, \citenamefont
  {Gordon}, \citenamefont {de~Cotret}, \citenamefont {Doiron-Leyraud},
  \citenamefont {Saito}, \citenamefont {Fukazawa}, \citenamefont {Kohori},
  \citenamefont {Kihou}, \citenamefont {Lee}, \citenamefont {Iyo},
  \citenamefont {Eisaki}, \citenamefont {Prozorov},\ and\ \citenamefont
  {Taillefer}}]{Reid2012PRL}%
  \BibitemOpen
  \bibfield  {author} {\bibinfo {author} {\bibfnamefont {J.~P.}\ \bibnamefont
  {Reid}}, \bibinfo {author} {\bibfnamefont {M.~A.}\ \bibnamefont {Tanatar}},
  \bibinfo {author} {\bibfnamefont {A.}~\bibnamefont {Juneau-Fecteau}},
  \bibinfo {author} {\bibfnamefont {R.~T.}\ \bibnamefont {Gordon}}, \bibinfo
  {author} {\bibfnamefont {S.~R.}\ \bibnamefont {de~Cotret}}, \bibinfo {author}
  {\bibfnamefont {N.}~\bibnamefont {Doiron-Leyraud}}, \bibinfo {author}
  {\bibfnamefont {T.}~\bibnamefont {Saito}}, \bibinfo {author} {\bibfnamefont
  {H.}~\bibnamefont {Fukazawa}}, \bibinfo {author} {\bibfnamefont
  {Y.}~\bibnamefont {Kohori}}, \bibinfo {author} {\bibfnamefont
  {K.}~\bibnamefont {Kihou}}, \bibinfo {author} {\bibfnamefont {C.~H.}\
  \bibnamefont {Lee}}, \bibinfo {author} {\bibfnamefont {A.}~\bibnamefont
  {Iyo}}, \bibinfo {author} {\bibfnamefont {H.}~\bibnamefont {Eisaki}},
  \bibinfo {author} {\bibfnamefont {R.}~\bibnamefont {Prozorov}}, \ and\
  \bibinfo {author} {\bibfnamefont {L.}~\bibnamefont {Taillefer}},\ }\bibfield
  {title} {\enquote {\bibinfo {title} {Universal heat conduction in the iron
  arsenide superconductor {K}{F}e$_2${A}s$_2$: Evidence of a $d$-wave state},}\
  }\href {\doibase 10.1103/PhysRevLett.109.087001} {\bibfield  {journal}
  {\bibinfo  {journal} {Phys. Rev. Lett.}\ }\textbf {\bibinfo {volume} {109}},\
  \bibinfo {pages} {087001} (\bibinfo {year} {2012})}\BibitemShut {NoStop}%
\bibitem [{\citenamefont {Okazaki}\ \emph {et~al.}(2012)\citenamefont
  {Okazaki}, \citenamefont {Ota}, \citenamefont {Kotani}, \citenamefont
  {Malaeb}, \citenamefont {Ishida}, \citenamefont {Shimojima}, \citenamefont
  {Kiss}, \citenamefont {Watanabe}, \citenamefont {Chen}, \citenamefont
  {Kihou}, \citenamefont {Lee}, \citenamefont {Iyo}, \citenamefont {Eisaki},
  \citenamefont {Saito}, \citenamefont {Fukazawa}, \citenamefont {Kohori},
  \citenamefont {Hashimoto}, \citenamefont {Shibauchi}, \citenamefont
  {Matsuda}, \citenamefont {Ikeda}, \citenamefont {Miyahara}, \citenamefont
  {Arita}, \citenamefont {Chainani},\ and\ \citenamefont {Shin}}]{Okazaki1314}%
  \BibitemOpen
  \bibfield  {author} {\bibinfo {author} {\bibfnamefont {K.}~\bibnamefont
  {Okazaki}}, \bibinfo {author} {\bibfnamefont {Y.}~\bibnamefont {Ota}},
  \bibinfo {author} {\bibfnamefont {Y.}~\bibnamefont {Kotani}}, \bibinfo
  {author} {\bibfnamefont {W.}~\bibnamefont {Malaeb}}, \bibinfo {author}
  {\bibfnamefont {Y.}~\bibnamefont {Ishida}}, \bibinfo {author} {\bibfnamefont
  {T.}~\bibnamefont {Shimojima}}, \bibinfo {author} {\bibfnamefont
  {T.}~\bibnamefont {Kiss}}, \bibinfo {author} {\bibfnamefont {S.}~\bibnamefont
  {Watanabe}}, \bibinfo {author} {\bibfnamefont {C.-T.}\ \bibnamefont {Chen}},
  \bibinfo {author} {\bibfnamefont {K.}~\bibnamefont {Kihou}}, \bibinfo
  {author} {\bibfnamefont {C.~H.}\ \bibnamefont {Lee}}, \bibinfo {author}
  {\bibfnamefont {A.}~\bibnamefont {Iyo}}, \bibinfo {author} {\bibfnamefont
  {H.}~\bibnamefont {Eisaki}}, \bibinfo {author} {\bibfnamefont
  {T.}~\bibnamefont {Saito}}, \bibinfo {author} {\bibfnamefont
  {H.}~\bibnamefont {Fukazawa}}, \bibinfo {author} {\bibfnamefont
  {Y.}~\bibnamefont {Kohori}}, \bibinfo {author} {\bibfnamefont
  {K.}~\bibnamefont {Hashimoto}}, \bibinfo {author} {\bibfnamefont
  {T.}~\bibnamefont {Shibauchi}}, \bibinfo {author} {\bibfnamefont
  {Y.}~\bibnamefont {Matsuda}}, \bibinfo {author} {\bibfnamefont
  {H.}~\bibnamefont {Ikeda}}, \bibinfo {author} {\bibfnamefont
  {H.}~\bibnamefont {Miyahara}}, \bibinfo {author} {\bibfnamefont
  {R.}~\bibnamefont {Arita}}, \bibinfo {author} {\bibfnamefont
  {A.}~\bibnamefont {Chainani}}, \ and\ \bibinfo {author} {\bibfnamefont
  {S.}~\bibnamefont {Shin}},\ }\bibfield  {title} {\enquote {\bibinfo {title}
  {Octet-line node structure of superconducting order parameter in
  {K}{F}e$_2${A}s$_2$},}\ }\href {\doibase 10.1126/science.1222793} {\bibfield
  {journal} {\bibinfo  {journal} {Science}\ }\textbf {\bibinfo {volume}
  {337}},\ \bibinfo {pages} {1314} (\bibinfo {year} {2012})}\BibitemShut
  {NoStop}%
\bibitem [{\citenamefont {Lederer}\ \emph {et~al.}(2015)\citenamefont
  {Lederer}, \citenamefont {Schattner}, \citenamefont {Berg},\ and\
  \citenamefont {Kivelson}}]{Lederer2015PRL}%
  \BibitemOpen
  \bibfield  {author} {\bibinfo {author} {\bibfnamefont {S.}~\bibnamefont
  {Lederer}}, \bibinfo {author} {\bibfnamefont {Y.}~\bibnamefont {Schattner}},
  \bibinfo {author} {\bibfnamefont {E.}~\bibnamefont {Berg}}, \ and\ \bibinfo
  {author} {\bibfnamefont {S.~A.}\ \bibnamefont {Kivelson}},\ }\bibfield
  {title} {\enquote {\bibinfo {title} {Enhancement of superconductivity near a
  nematic quantum critical point},}\ }\href
  {http://link.aps.org/doi/10.1103/PhysRevLett.114.097001} {\bibfield
  {journal} {\bibinfo  {journal} {Phys. Rev. Lett.}\ }\textbf {\bibinfo
  {volume} {114}},\ \bibinfo {pages} {097001} (\bibinfo {year}
  {2015})}\BibitemShut {NoStop}%
\bibitem [{\citenamefont {Kang}\ and\ \citenamefont
  {Fernandes}(2016)}]{Kang2016PRL}%
  \BibitemOpen
  \bibfield  {author} {\bibinfo {author} {\bibfnamefont {J.}~\bibnamefont
  {Kang}}\ and\ \bibinfo {author} {\bibfnamefont {R.~M.}\ \bibnamefont
  {Fernandes}},\ }\bibfield  {title} {\enquote {\bibinfo {title}
  {Superconductivity in {F}e{S}e thin films driven by the interplay between
  nematic fluctuations and spin-orbit coupling},}\ }\href
  {http://link.aps.org/doi/10.1103/PhysRevLett.117.217003} {\bibfield
  {journal} {\bibinfo  {journal} {Phys. Rev. Lett.}\ }\textbf {\bibinfo
  {volume} {117}},\ \bibinfo {pages} {217003} (\bibinfo {year}
  {2016})}\BibitemShut {NoStop}%
\end{thebibliography}%

\end{document}